\documentclass[conference]{IEEEtran}
\IEEEoverridecommandlockouts
\usepackage{cite}
\usepackage{amsmath,amssymb,amsfonts}
\usepackage{algorithmic}
\usepackage{graphicx}
\usepackage{textcomp}
\usepackage{hyperref}
\usepackage{xcolor}
\usepackage[ruled,vlined]{algorithm2e}
\usepackage{graphicx,wrapfig,lipsum}
\usepackage{color}
\usepackage[a4paper, total={184mm,239mm}]{geometry}
\usepackage{lettrine}
\usepackage{comment}
\usepackage{booktabs,chemformula}
\usepackage{graphicx}
\usepackage{subfigure}
\usepackage{amsmath}
\usepackage{xcolor}
\usepackage{color, soul}
\usepackage{tabularx,booktabs}
\usepackage{multirow}
\usepackage{ulem}
\usepackage{tabularx}
\usepackage{array}
\usepackage{makecell}

\newcolumntype{L}[1]{>{\raggedright\arraybackslash}p{#1}}
 
\newcolumntype{C}[1]{>{\centering\arraybackslash}p{#1}}

\newcolumntype{C}{>{\centering\arraybackslash}X} 
\usepackage{lipsum} % filler text

\newcommand{\red}{\color{red}}

\def\BibTeX{{\rm B\kern-.05em{\sc i\kern-.025em b}\kern-.08em
    T\kern-.1667em\lower.7ex\hbox{E}\kern-.125emX}}

\hypersetup{hidelinks} % hides the cheap green boxes around citations

\begin{document}

%\title{Approximate JPEG {\red{\sout{Compression} Technique}} Hardware for Energy-Constrained Image Sensors}

\title{Approximate JPEG Compression Hardware for Energy-Constrained Image Sensors}
\title{Approximate JPEG Compression Technique for Energy-Constrained Image Sensors}
\title{Multiplier-less DCT and Approximate Quantization Techniques for JPEG Compression in Energy-Constrained Image Sensors}
\title{Multiplier-less DCT and Approximate Quantization Techniques for Energy-Constrained Image Sensors}
\title{Approximate DCT and Quantization Techniques for Energy-Constrained Image Sensors}

\author{
        Ming-Che Li*,~\IEEEmembership{Student~Member,~IEEE}, 
        Archisman Ghosh*,~\IEEEmembership{Student~Member,~IEEE},\\
        and~Shreyas~Sen,~\IEEEmembership{Senior~Member,~IEEE.} \\
        \newline *Authors contributed equally}

\maketitle

\begin{abstract}
Recent expansions in multimedia devices for many applications, such as surveillance, self-driving cars, and healthcare, gather enormous amounts of real-time images for processing and inference. The images are first compressed using compression schemes, like JPEG, before processing to reduce storage costs and additional power requirements for transmitting the captured data in this era of emerging ultra-wide-band communication and human-body communication.  The JPEG algorithm realizes image compression using simplistic matrix manipulations, making it preferable for hardware implementations. Furthermore, due to inherent error resilience and imperceptibility in images, JPEG can be approximated to reduce the required computation/processing power and area. This work demonstrates the first end-to-end approximation computing-based optimization of JPEG hardware using i) an approximate division realized using bit-shift operators to reduce the complexity of the computationally intensive quantization block, ii) loop perforation, and iii) precision scaling on top of a multiplier-less fast DCT architecture to achieve an extremely energy-efficient JPEG compression unit which will be a perfect fit for power/bandwidth-limited scenario. Furthermore, a gradient descent-based heuristic composed of two conventional approximation strategies, i.e., Precision Scaling and Loop Perforation, is implemented for tuning the degree of approximation to trade off energy consumption with the quality degradation of the decoded image. 
The entire RTL (Register-Transfer Level) design is coded in Verilog HDL, synthesized using the industry-standard tool, and mapped to TSMC 65nm CMOS technology. and simulated using Cadence Spectre Simulator under 25$^{\circ}$\textbf{C}, TT (Typical/Typical) corner. The approximate division approach in the quantization block achieved around $\textbf{28\%}$ reduction in the active design area.
The heuristic-based approximation technique combined with accelerator optimization achieves a significant energy reduction of $\textbf{36\%}$ for a minimal image quality degradation of $\textbf{2\%}$ SAD (Sum of Absolute Difference). Simulation results also show that the proposed architecture consumes 15uW at the DCT and quantization stages to compress a colored 480p image at 6fps.

\end{abstract}

\begin{IEEEkeywords}
Approximate computing, approximate divider, precision scaling, loop perforation, energy-efficient, image sensor, in-sensor analytics
\end{IEEEkeywords}

%%%%%%%%%%%%%%%%%%%%%%%%%%%%%%%%% INTRODUCTION %%%%%%%%%%%%%%%%%%%%%%%%%%%%%%%%% 
\section{Introduction}
\begin{figure}[tbp]
\centerline{\includegraphics[width=0.38\textwidth]{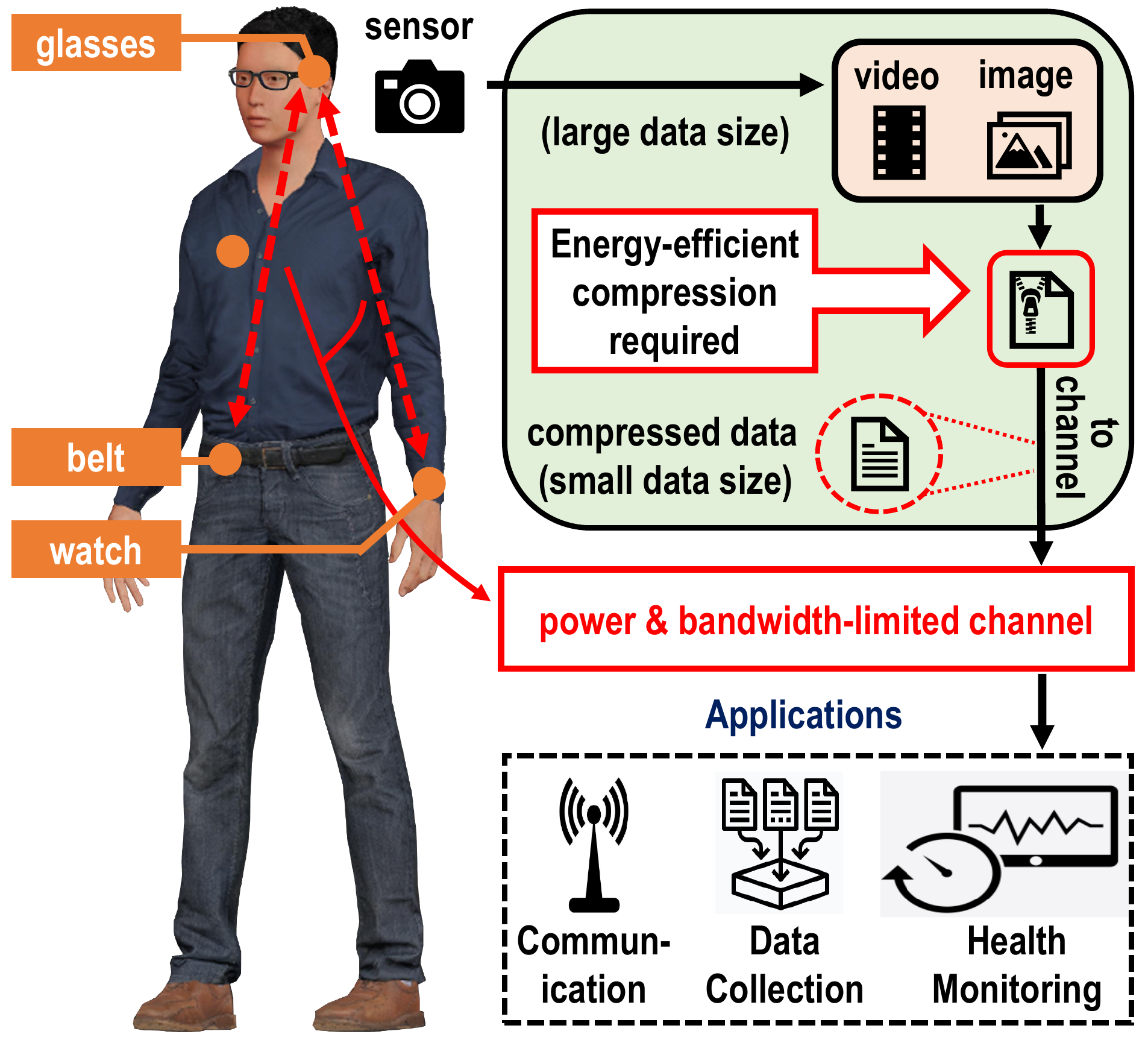}}
\caption{Significance of energy-constrained compression hardware in  a power \& bandwidth-limited scenario.}
\label{fig:introduction}
%\vspace{-0.1in}
\end{figure}

\lettrine[lines=2]{U}{sage} of multimedia devices has expanded exponentially in recent years in every application, such as surveillance, self-driving cars, and healthcare. These sensors generate enormous amounts of data in images or videos that require processing and storage.  As these sensors operate in an energy-constrained environment, as shown in Fig.~\ref{fig:introduction}, the collected images are often compressed at the source using conventional image compression techniques to mitigate the energy expended in transmission, processing, and storage.

Image compression algorithms can be categorized into two classes based on the error they introduced, i.e., lossy, and lossless compressions. Lossless compressions, like PNG (Portable Network Graphics) and TIFF (Tagged Image File Format), ensure
perfect reconstruction of the image at the cost of reduced compression (usually 2:1 \cite{9178690}). On the other hand, lossy schemes such as JPEG (Joint Photographics Experts Group), PGF (Progressive Graphics File), and JPEG2000 \cite{4908583} provide higher compression while seeking to reconstruct a visually similar image with imperceptible degradation in quality. 

Among these lossy compression schemes, JPEG is one of the most common image compression methods utilized in energy-constrained edge devices due to its lightweight nature and broad compatibility. Although JPEG2000 provides better compression efficiency and less image quality degradation at lower bit rates because of its compression algorithm\cite{The_JPEG_2000_still_image_compression_standard} \cite{The_JPEG2000_still_image_coding_system_an_overview}\cite{JPEGvsJPEG2000_an_objective_comparison_of_image_encoding_quality}\cite{Image_Quality_Comparison_Between_JPEG_and_JPEG2000_I_Psychophysical_Investigation}, JPEG has gained popularity over JPEG2000 due to its lower hardware requirements, which is a more critical factor for performing image compression in energy-constrained environments.

\begin{figure}[!ht]
% \centerline{\includegraphics[scale=0.35]{Metal.png}}
%\centerline{\includegraphics[width=0.48\textwidth]{distribution.png}}
\centerline{\includegraphics[width=0.48\textwidth]{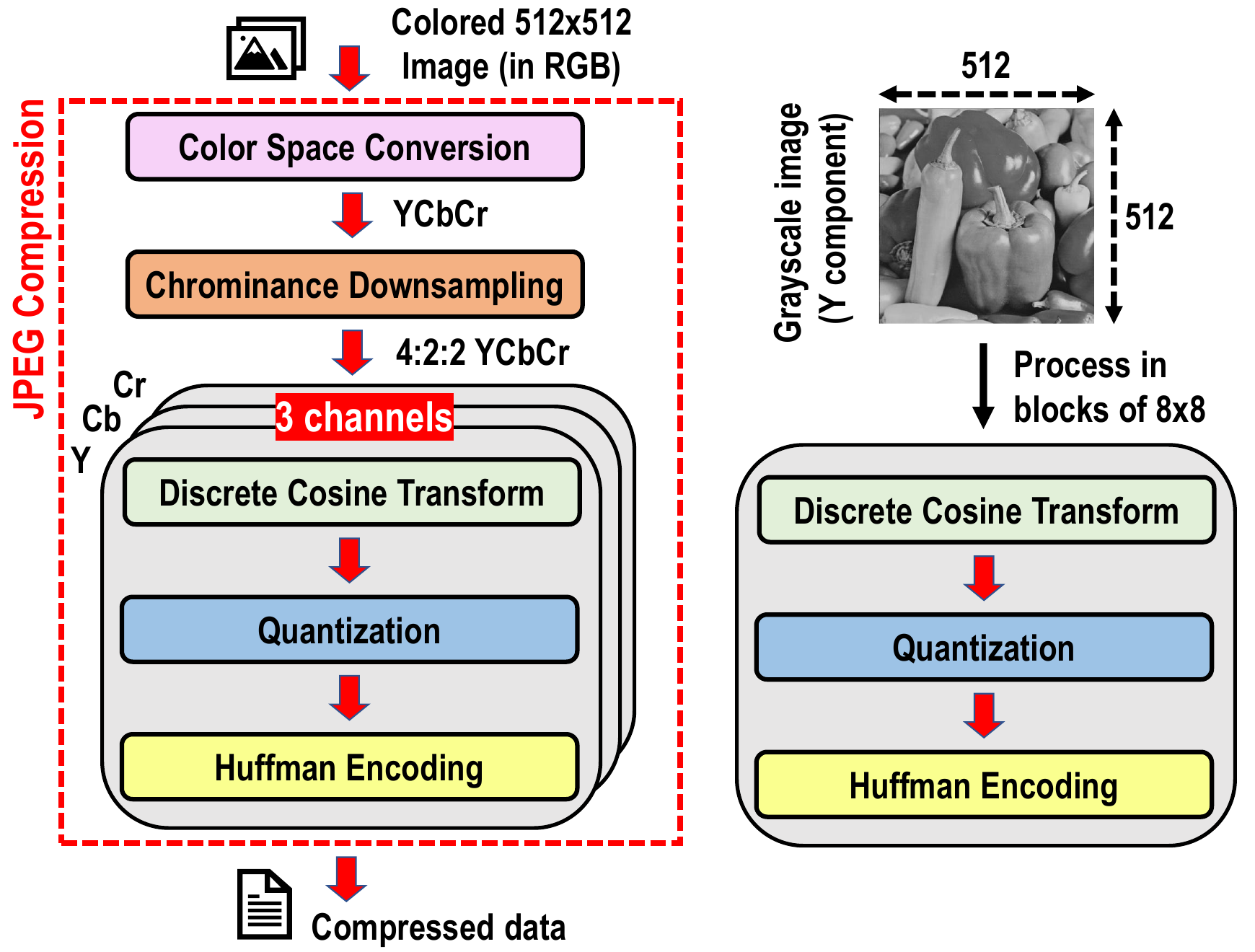}}
\caption{Flowchart of standard JPEG compression.}
\label{fig:jpeg_flowchart}
%\vspace{-0.1in}
\end{figure}
To date, approximate solutions for JPEG image compression are concentrated on algorithm or hardware levels. Previous works \cite{BAS08, BAS09, BAS11, ANaE8DCT} present mathematical approximations on the Discrete Cosine Transform (DCT) matrix in which fractional coefficients are replaced with integers or negative powers of two, but no corresponding implementation on hardware is proposed in the literature. On the other hand, hardware-target approximation literature, such as \cite{dac_2016_sachin} (performing bit truncation at DCT stage), \cite{TCAD_anand}\cite{tvlsi_kaushik}\cite{Haider18} (using approximate adders), or \cite{date_2019_cade} (replacing the quantization using approximate divider), only focus on the single stage optimization. They fail to produce an energy-optimal compression circuit that performs approximate computing on more than one of the JPEG compression stages.

In this work, we present an accelerated JPEG computing unit that incorporates an approximate quantization block and two well-known approximation techniques—loop perforation and precision scaling—on top of a multiplier-less fast DCT architecture to create a solution suitable for energy-constrained devices. A lightweight heuristic is also developed to estimate and fine-tune the degree of loop perforation and precision scaling without introducing significant degradation in the output image quality.

\par
In brief, the proposed work has three-fold contributions: 

1) We propose an approximate low-power area-efficient JPEG compression architecture focused on restructuring the computation-heavy quantization block. No such hardware-efficient image compression circuit has been reported that utilizes the bit shift operators-based division modules for optimizing the same to date. \par

2) A gradient descent-based heuristic is proposed that employs the commonly used loop perforation and precision scaling strategies to automatically configure the degree of approximation in the compression engine to reduce the energy required for processing while maintaining the quality of the decoded image within acceptable limits. \par

3) The reconfigurable architecture with optimized division modules and the implemented heuristics achieves a significant energy reduction of $36\%$ for a minimal image quality degradation of SAD ($2\%$).  \par 
The overall implementation achieves \textcolor{black}{28\% area improvement with respect to the baseline design.} It is important to note that our baseline design is already optimized using multiplier-less DCT architecture. Simulation results show that the proposed architecture consumes only 15uW power while compressing 480p images at a 6fps rate.
%Note that the reported area and power values are measured in the post-synthesis simulation by Cadence Virtuoso Simulator.

The rest of the paper is organized as follows. The JPEG compression technique is briefly discussed in Section~\ref{sec:JPEG_compression}. Section~\ref{sec:approx_tech} dives into the implemented approximation techniques along with the proposed heuristics that dynamically selects the optimum approximation knobs for operation. Simulation methodology is discussed in Section~\ref{sec:sim_setup}. Finally, we discuss the results in Section~\ref{sec:sim_results}, before concluding the paper in Section~\ref{sec:final}. 

\begin{figure}[!ht]
% \centerline{\includegraphics[scale=0.35]{Metal.png}}
%\centerline{\includegraphics[width=0.48\textwidth]{distribution.png}}
\centerline{\includegraphics[width=0.4\textwidth]{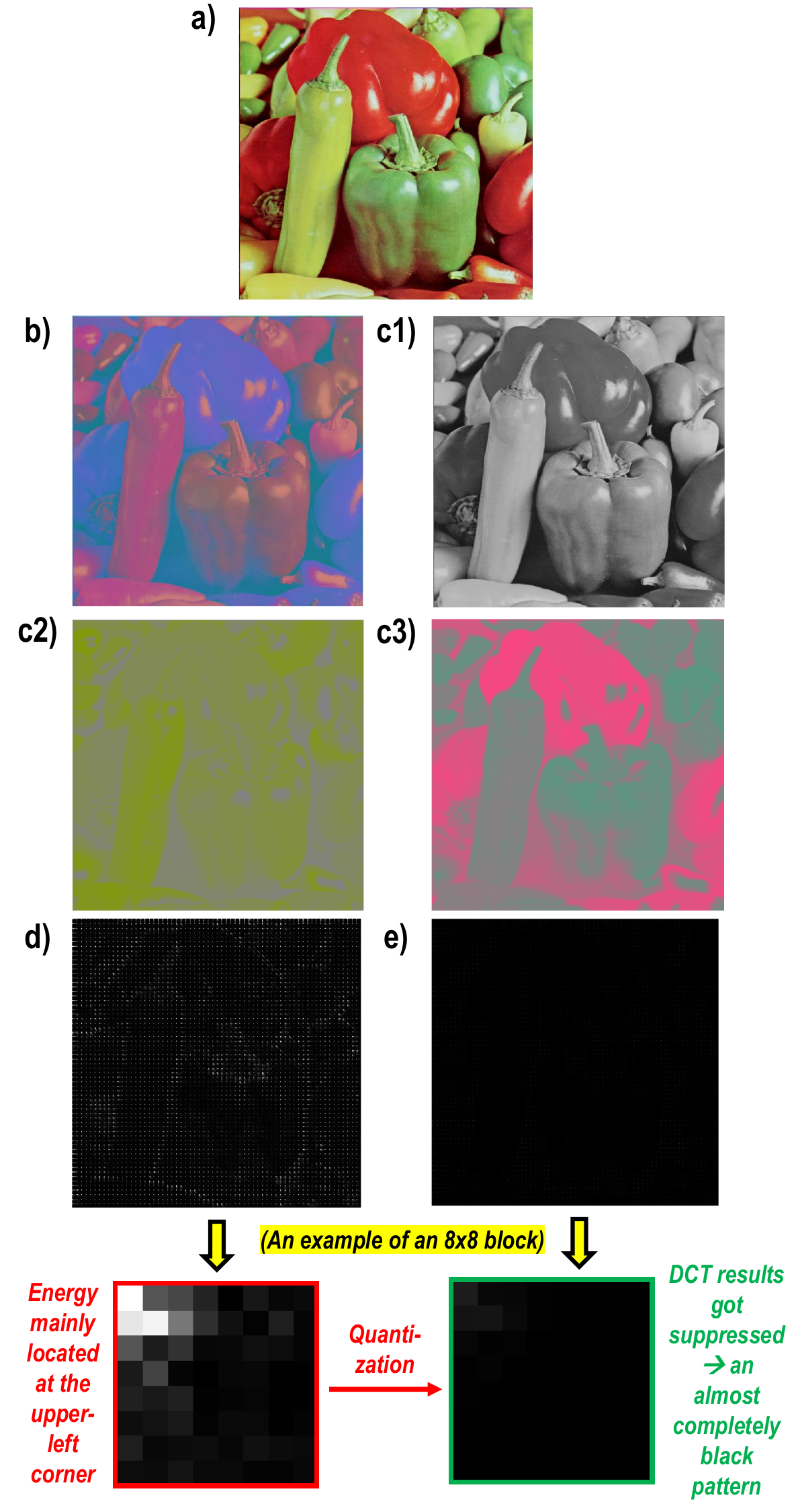}}
\caption{A visual example of the JPEG compression process: (a) colored (RGB format) image to be compressed. (b) colored image in YCbCr format. (c1) Y component represents the brightness of the original image. (c2)-(c3) Cb and Cr components represent the strength of blue and red signals of the original image, respectively. (d) the DCT (given by Eq. (1)) result of the Y component. (e) quantized result of (d).}
\label{fig:jpeg_compression_images_different_stages}
\vspace{-0.1in}
\end{figure}
%%%%%%%%%%%%%%%%%%%%%%%%%%%%%%%%% II. Image Compression %%%%%%%%%%%%%%%%%%%%%%%%%%%%%%%%% 
\section{Image Compression Technique}\label{sec:JPEG_compression}

% TODO: Ask if there is a good reference or if a reference is needed for the JPEG algorithm
% Link: 

The Joint Photographics Experts Group developed the ISO (International Organization for Standardization) for the standard JPEG image compression \cite{125072} algorithm, which can be broken down into five essential stages. To compress one 512x512 pixels colored image, the image has to be run through the following stages, as shown in Fig.~\ref{fig:jpeg_flowchart}: 

\begin{itemize}
    \item Stage 1: Convert pixel values from RGB to YCbCr format. (Different from RGB color space, YCbCr is another color space where a colored image is described in terms of brightness (Y) and the chroma strength of blue (Cb) and red (Cr) signals.)
    \item Stage 2: Downsample the chrominance component (Cb and Cr) for every 2x2 pixel block.
    \item Stage 3: Discrete Cosine Transform (DCT) on 8x8 pixel block.
    \item Stage 4: Quantize DCT output matrix via element-wise division by quantization matrix.
    \item Stage 5: Perform Huffman Encoding on the zigzag traversal of the quantized matrix to keep low-frequency components at the beginning of the serial chain and insignificant high-frequency components at the end.
\end{itemize}

%Taking advantage of the fact that the human eye is relatively insensitive to chromaticity compared to luminance, the first two stages of JPEG compression appropriately reduce the size of the data used to represent the color intensity without affecting the visual quality of the image. 
JPEG compression exploits the fact that the human eye is relatively insensitive to (1) chromaticity compared to luminosity and (2) high-frequency spatial changes in an image to compress an image. Because of property (1), the size of the data used to represent the color intensity of an image can be appropriately reduced without affecting the image's visual quality. This data size reduction is realized by the first two stages of JPEG compression, where an image is converted into the space of luminance and chrominance first, and its chrominance components are then sampled every 2 pixels in the horizontal direction while every luminance pixel is retained. On the other hand, high-frequency spatial changes in an image are suppressed in  DCT, quantization, and Huffman encoding three stages. Note that starting from Stage 3, the compression is performed in the unit of 8x8 pixel blocks; moreover, three channels are required to process the compression of the original image's Y, Cb, and Cr components. Fig.~\ref{fig:jpeg_compression_images_different_stages} demonstrates the visual example of how an image is compressed.
\par
After the first two stages, the input image is processed into chunks of pixel blocks. One block is represented as an 8x8 matrix $\mathbf{M}$. The \textbf{Discrete Cosine Transform} on $\mathbf{M}$ is given by the matrix multiplication: 

\begin{equation}
\mathbf{D} = \mathbf{TMT}^{'}
\end{equation}

where $\mathbf{D}$ is the DCT of $\mathbf{M}$, $\mathbf{T}$ is the transformation matrix, and $\mathbf{T}^{'}$ is transpose of $\mathbf{T}$. The transformation matrix $\mathbf{T}$ is defined as: 

\begin{equation}
\label{eq:DCT_matrix}
{t}_{ij} =
    \begin{cases}
        \frac{1}{\sqrt{8}}, & \text{if } i = 0 \\
        \sqrt{\frac{2}{8}} \cdot \cos{\frac{(2j + 1)i\pi{}}{16}}, & \text{if } i > 0
    \end{cases}
\end{equation}

where ${t}_{ij}$ is the element in $\mathbf{T}$ and subscripts $i$, $j$ are from 0 to 7 and represent the row position and the column position, respectively.

\begin{figure}[tbp]
% \centerline{\includegraphics[scale=0.35]{Metal.png}}
%\centerline{\includegraphics[width=0.48\textwidth]{distribution.png}}
\centerline{\includegraphics[width=0.45\textwidth]{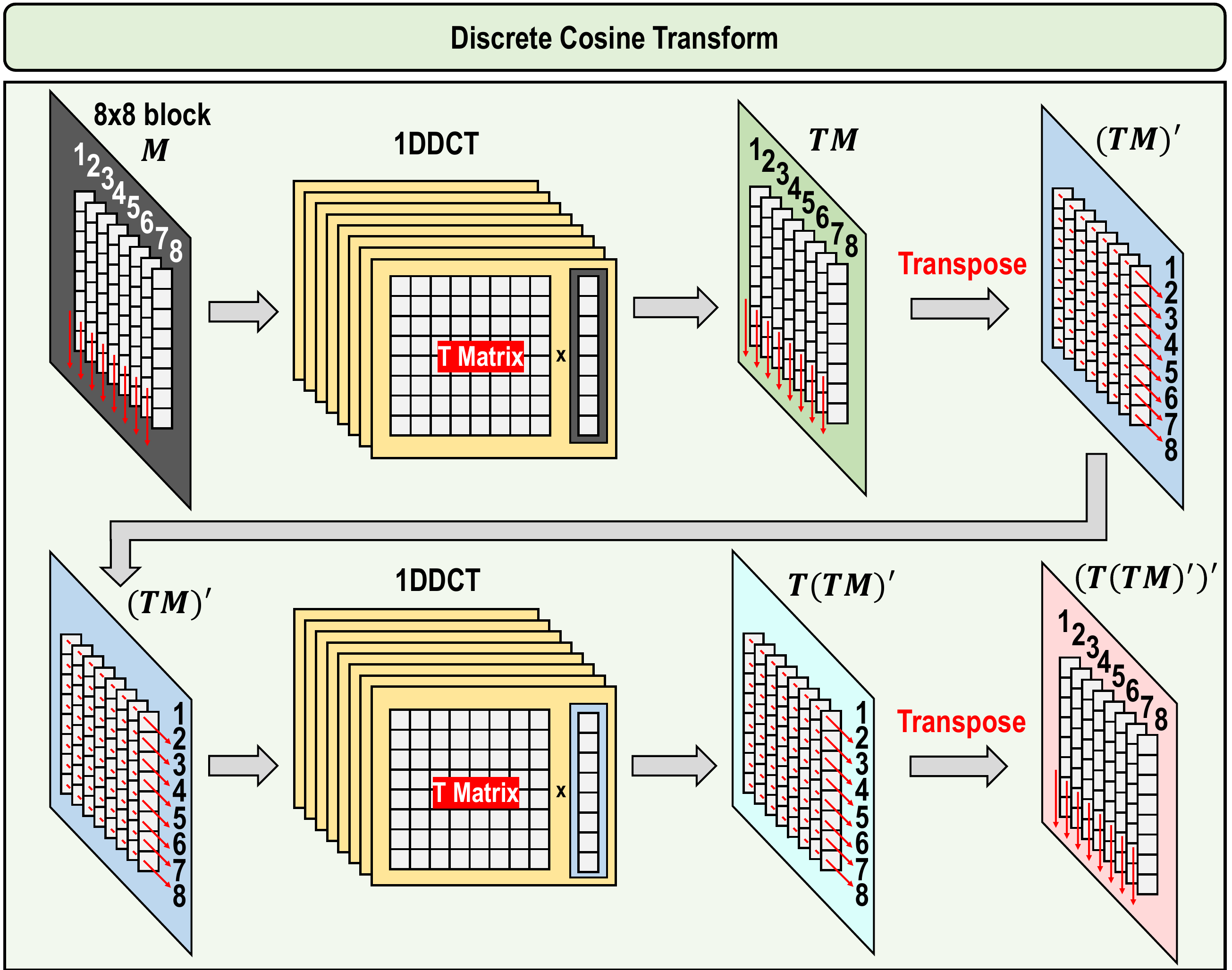}}
%\caption{From 2D-DCT to 2 rounds of 1D-DCT.}
\caption{Dividing 2-D DCT into 1-D DCT.}
\label{fig:two_rounds_of_1DDCT}
\vspace{-0.1in}
\end{figure}

\begin{figure}[tbp]
% \centerline{\includegraphics[scale=0.35]{Metal.png}}
%\centerline{\includegraphics[width=0.48\textwidth]{distribution.png}}
\centerline{\includegraphics[width=0.5\textwidth]{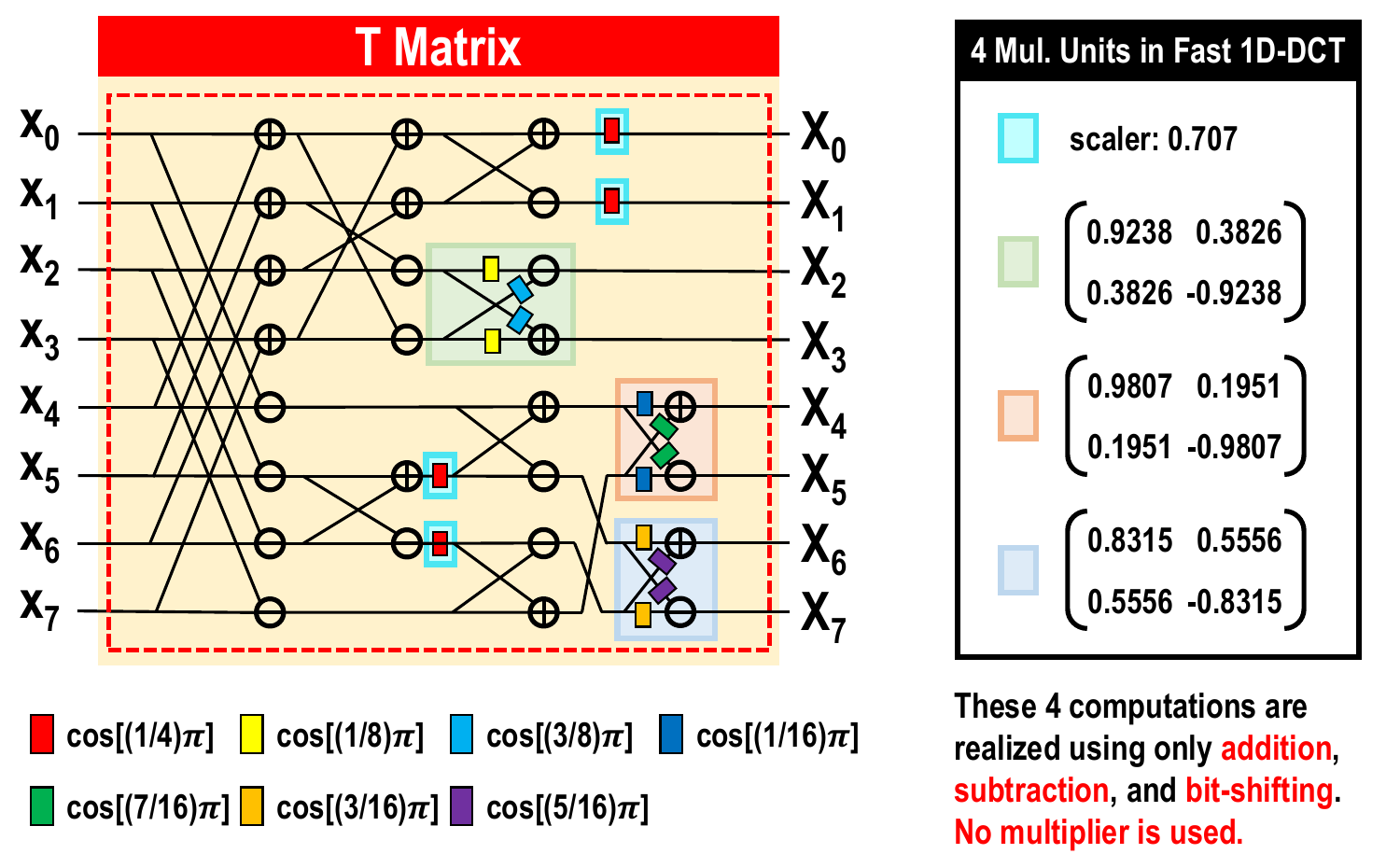}}
\caption{1D-FDCT: Butterfly diagram for 8-point 1D-DCT. Each colored small block represents a multiplication with a specific cosine factor.}
\label{fig:1DFDCT}
\vspace{-0.1in}
\end{figure}

\begin{figure}[tbp]
% \centerline{\includegraphics[scale=0.35]{Metal.png}}
%\centerline{\includegraphics[width=0.48\textwidth]{distribution.png}}
\centerline{\includegraphics[width=0.5\textwidth]{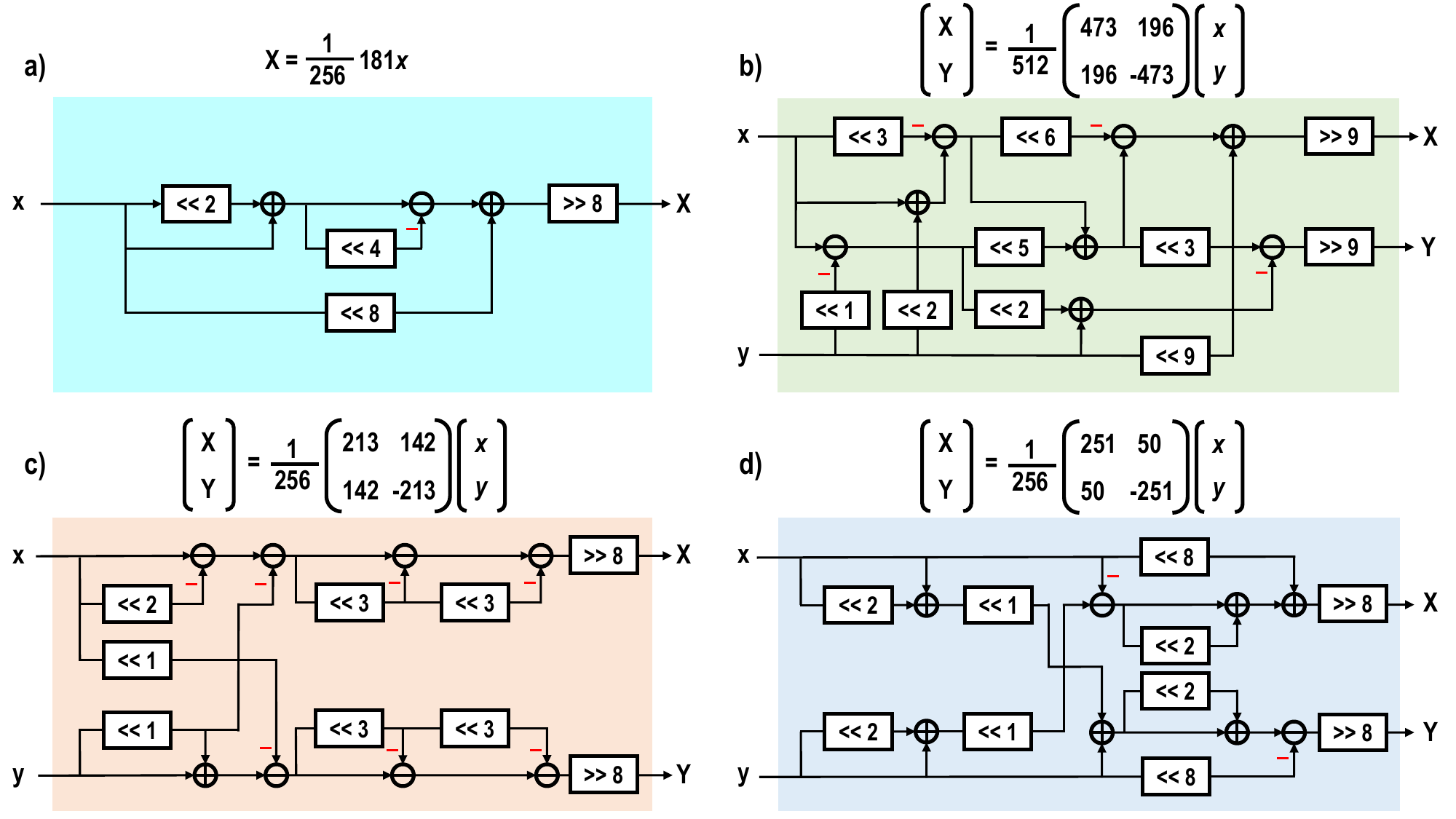}}
\caption{Detailed multiplier-less implementation of (a) the scaling operation and (b)-(d) three 2x2 matrix multiplications in Fig. \ref{fig:1DFDCT}.}
\label{fig:1DFDCT_4butterfly}
\vspace{-0.1in}
\end{figure}

In this work, to build an energy-efficient JPEG compression circuit, we implement Eq. (1) not by the direct matrix multiplication method but by the \textbf {fast DCT (FDCT)} method \cite{1093941}. Since Eq. (1) can be written as:
\begin{equation}
    \mathbf{TMT'} = \mathbf{[(T)(TM)']'},
\end{equation}
the computation of $\mathbf{D}$ can be treated as 2 rounds of one-dimentional DCT (1D-DCT), as shown in Fig.~\ref{fig:two_rounds_of_1DDCT}. The 1D-DCT is the multiplication of the transformation matrix $\mathbf{T}$ and any given 8x1 vector $\mathbf{x}$. The computation of 1D-DCT can be mathematically 'accelerated' using the fast-architecture. The fast architecture with different types of butterfly units is shown in Fig.~\ref{fig:1DFDCT}, which reduces the multiplications of  $\mathbf{Tx}$ from 64 to 16. On the hardware level, an energy-efficient 1D-FDCT unit is achieved by adopting the multiplier-less approach\cite{5156873}, where four types of multiplications in 1D-FDCT (one scaler multiplication and three 2x2 matrix multiplications) are implemented by adders and shifters. Eq. (\ref{eq:1DDCT_scaler_mathform})(\ref{eq:1DDCT_Butterfly1_mathform})(\ref{eq:1DDCT_Butterfly2_mathform})(\ref{eq:1DDCT_Butterfly3_mathform}) delineate the precise mathematical expression for the four multiplications, with $x$ and $y$ representing the inputs and $X$ and $Y$ denoting the multiplication outputs. Eq. (\ref{eq:1DDCT_scaler_shiftaddform})(\ref{eq:1DDCT_Butterfly1_shiftaddform})(\ref{eq:1DDCT_Butterfly2_shiftaddform})(\ref{eq:1DDCT_Butterfly3_shiftaddform}) explicitly illustrate how these multiplications are executed through the shift add method. Fig.~\ref{fig:1DFDCT_4butterfly} portrays the associated hardware-level implementation, comprising solely adders, subtractors, and barrel-shifters. Our DCT core does not employ any general multiplier.

$Scaler \; multiplication$:
\begin{subequations}
    \label{eq:1DDCT_scaler}
    \begin{equation}
        \label{eq:1DDCT_scaler_mathform}
        X = 0.707x \backsimeq \frac{181}{256}x = \frac{5(1-16)+256}{256}x
    \end{equation}
    \begin{equation}
      \label{eq:1DDCT_scaler_shiftaddform}
      \begin{aligned}
         X &= b>>8 \\        
         &\quad b = (a - a<<4) + (x<<8) \\
         &\quad\quad a = x + (x<<2)
      \end{aligned}
    \end{equation}
\end{subequations}
$Butterfly \; matrix \; multiplication \; (i)$:
\begin{subequations}
\label{eq:1DDCT_Butterfly1}
    \begin{equation}
        \label{eq:1DDCT_Butterfly1_mathform}
        \begin{bmatrix} X \\ Y \end{bmatrix} = 
        \begin{bmatrix} 0.9238&0.3836 \\ 0.3836&-0.9238 \end{bmatrix}
        \begin{bmatrix} x \\ y \end{bmatrix}
        \backsimeq \frac{1}{2^{9}} \begin{bmatrix} 473&196 \\ 196&-473 \end{bmatrix}
        \begin{bmatrix} x \\ y \end{bmatrix}
    \end{equation}  
    \begin{equation}
      \label{eq:1DDCT_Butterfly1_shiftaddform}
      \begin{aligned}
        X &= \frac{1}{512}(473x+196y) = d_{x} >> 9\\        
        Y &= \frac{1}{512}(196x-473y) = d_{y} >> 9\\
        &\quad d_{x} = c_{xy} - (b_{x}<<6) + (y<<9)\\
        &\quad d_{y} = (c_{xy}<<3) - b_{y}\\
        &\quad\quad c_{xy} = b_{x} + (a_{xy}<<5)\\
        &\quad\quad b_{x} = x - (x<<3) + (y<<2)\\
        &\quad\quad b_{y} = (a_{xy}<<2) +y\\
        &\quad\quad\quad a_{xy} = x - (y<<1)
      \end{aligned}
    \end{equation}
\end{subequations}
$Butterfly \; matrix \; multiplication \; (ii)$:
\begin{subequations}
\label{eq:1DDCT_Butterfly2}
  \begin{equation}
    \label{eq:1DDCT_Butterfly2_mathform}
    \begin{bmatrix} X \\ Y \end{bmatrix} = 
    \begin{bmatrix} 0.9807&0.1951 \\ 0.1951&-0.9807 \end{bmatrix} 
    \begin{bmatrix} x \\ y \end{bmatrix}
    \backsimeq \frac{1}{2^{8}} \begin{bmatrix} 213&142 \\ 142&-213 \end{bmatrix}
    \begin{bmatrix} x \\ y \end{bmatrix}
 \end{equation}
 \begin{equation}
  \label{eq:1DDCT_Butterfly2_shiftaddform}
  \begin{aligned}
    X &= \frac{1}{256}(213x+142y) = b_{x} >> 8\\        
    Y &= \frac{1}{256}(142x-213y) = b_{y} >> 8\\
    &\quad b_{x} = -71a_{x}\\
    &\quad b_{y} = -71a_{y}\\
    &\quad\quad a_{x} = x - (x<<2) - (y<<1)\\
    &\quad\quad a_{y} = -(x<<1) + y + (y<<1)
  \end{aligned}
\end{equation}
\end{subequations}
$Butterfly \; matrix \; multiplication \; (iii)$:
\begin{subequations}
\label{eq:1DDCT_Butterfly3}
  \begin{equation}
    \label{eq:1DDCT_Butterfly3_mathform}
    \begin{bmatrix} X \\ Y \end{bmatrix} = 
    \begin{bmatrix} 0.8315&0.5556 \\ 0.5556&-0.8315 \end{bmatrix} 
    \begin{bmatrix} x \\ y \end{bmatrix}
    \backsimeq \frac{1}{2^{8}} \begin{bmatrix} 251&50 \\ 50&-251 \end{bmatrix}
    \begin{bmatrix} x \\ y \end{bmatrix}
\end{equation}
 \begin{equation}
  \label{eq:1DDCT_Butterfly3_shiftaddform}
  \begin{aligned}
    X &= \frac{1}{256}(251x+50y) = c_{x} >> 8\\        
    Y &= \frac{1}{256}(50x-251y) = c_{y} >> 8\\
    &\quad c_{x} = (x<<8) + b_{x}\\
    &\quad c_{y} = -(y<<8) + b_{y}\\
    &\quad\quad b_{x} = a_{x} + (a_{x}<<2) \\
    &\quad\quad b_{y} = a_{y} + (a_{y}<<2) \\
    &\quad\quad\quad a_{x} = -x +  ((y + (y<<2))>>2)\\
    &\quad\quad\quad a_{y} = ((x + (x<<2))>>2) + y 
  \end{aligned}
\end{equation} 
\end{subequations}

\begin{comment}
\begin{equation}
  \label{eq:t}
  \begin{aligned}
    X &= \frac{213x+142y}{512} = \frac{(-71) (-3x-2y)}{2^{8}}\\        
    Y &= \frac{142x-213y}{512} = \frac{(-71) (-2x+3y)}{2^{8}}
  \end{aligned}
\end{equation}
\end{comment}

\begin{comment}
\begin{equation}
  \label{eq:t}
  \begin{aligned}
    X &= \frac{251x+50y}{256} = \frac{256x+5(-x-10y)}{256}\\        
    Y &= \frac{50x-251y}{256} = \frac{5(10x+y)-256y}{256}
  \end{aligned}
\end{equation}    
\end{comment}

\begin{comment}
    \begin{equation}
      \label{eq:t}
      \begin{aligned}
        X &= \frac{251x+50y}{256} = c_{x} >> 8\\        
        Y &= \frac{50x-251y}{256} = c_{y} >> 8\\
        c_{x} &= (x<<8) + b_{x}\\
        c_{y} &= -(y<<8) + b_{y}\\
        b_{x} &= a_{x} + (a_{x}<<2) \\
        b_{y} &= a_{y} + (a_{y}<<2) \\
        a_{x} &= -x +  ((y + (y<<2))>>2)\\
        a_{y} &= ((x + (x<<2))>>2) + y 
      \end{aligned}
    \end{equation}
\end{comment}

\par After the 8x8 pixel block is converted into the frequency domain, the DCT matrix $\mathbf{D}$ is next quantized by a designated quantization matrix $\mathbf{Q}$. The quantization stage generally performs element-wise division by directly instantiating divider blocks. The user can choose different quantization matrices based on the trade-off between the image quality and the compression level. The higher the quality level a $\mathbf{Q}$ matrix has, the less image compression will be. In other words, the information of the original image will not be discarded much, so when the image is reconstructed, higher image quality can be obtained. Typically, the quantization matrix with a quality level of 50, $\mathbf{Q_{50}}$, is used at this stage to achieve a good decompressed image quality and a decent compression ratio. \begin{equation}
    \mathbf{Q_{50}} = 
    \begin{bmatrix}
    \label{eq:Q50_unmodified}
    16 & 11 & 10 & 16 & 24 & 40 & 51 & 61 \\
    12 & 12 & 14 & 19 & 26 & 58 & 60 & 55 \\
    14 & 13 & 16 & 24 & 40 & 57 & 69 & 56 \\
    14 & 17 & 22 & 29 & 51 & 87 & 80 & 62 \\
    18 & 22 & 37 & 56 & 68 & 109 & 103 & 77 \\
    24 & 35 & 55 & 64 & 81 & 104 & 113 & 92 \\
    49 & 64 & 78 & 87 & 103 & 121 & 120 & 101 \\
    72 & 92 & 95 & 98 & 112 & 100 & 103 & 99 \\
    \end{bmatrix}
\end{equation}
If a higher image quality is needed, a quantization matrix with a quality level greater than 50 is necessitated. The required matrix is obtained by multiplying $\mathbf{Q_{50}}$ with a scaling factor of $(100-quality \; level)/50$ and then rounded and clipped so that all entry values are integers ranging from 1 to 255. For example, $\mathbf{Q_{90}}$, the quantization matrix with a quality level of 90, is given by
 \begin{equation}
    \mathbf{Q_{90}} = 
    \begin{bmatrix}
    \label{eq:Q90_unmodified}
    3 & 2 & 2 & 3 & 5 & 8 & 10 & 12 \\
    2 & 2 & 3 & 4 & 5 & 12 & 12 & 11 \\
    3 & 3 & 3 & 5 & 8 & 11 & 14 & 11 \\
    3 & 3 & 4 & 6 & 10 & 17 & 16 & 12 \\
    4 & 4 & 7 & 11 & 14 & 22 & 21 & 15 \\
    5 & 7 & 11 & 13 & 16 & 12 & 23 & 18 \\
    10 & 13 & 16 & 17 & 21 & 24 & 24 & 21 \\
    14 & 18 & 19 & 20 & 22 & 20 & 20 & 20 \\
    \end{bmatrix}.
\end{equation}

The output of this stage is given by an 8x8 matrix $\mathbf{C}$, with $\mathbf{C} = \mathbf{D} ~./~ \mathbf{Q}$, where $./$ represents the element-wise division operator. (We note that the entry values of $\mathbf{Q_{90}}$ are smaller than those of $\mathbf{Q_{50}}$, which implies $\mathbf{Q_{90}}$ will result in a $\mathbf{C}$ with larger entry values than $\mathbf{Q_{50}}$ will. That is, there will be more information for the image's reconstruction and higher reconstructed image quality can be obtained.) 
{\red} 

\par
%Finally, the last stage performs the \textbf{Huffman Encoding} process on $\mathbf{C}$ by zig-zag traversal starting from the element in the top left ($\mathbf{C}_{0,0}$) to the bottom right ($\mathbf{C}_{7,7}$). After concatenating the bits in this traversal as an input bit stream, a binary tree of the most common N-bit groups is created where a traversal of the encoding tree in the left or right directions maps to a binary bit (0 or 1). This allows an N-bit chunk of zeros (most common) to map to only a single bit after Huffman Encoding. After encoding, a compressed bit stream is created and sent to the communication channel.%
Finally, the quantized result is compressed into a bit stream using Huffman Encoding. The elements $\mathbf{C}$ are concatenated into a 1D vector based on the zig-zag traversal. After concatenating the bits in this traversal as an input bit stream, a binary tree of the most common N-bit groups is created where a traversal of the encoding tree in the left or right directions maps to a binary bit (0 or 1). This allows an N-bit chunk of zeros (most common) to map to only a single bit after Huffman Encoding. After encoding, a compressed bit stream is created and sent to the communication channel.

\section{Approximation Techniques}\label{sec:approx_tech}
To date, the approximate JPEG compression technique has been explored by using only bit truncation \cite{dac_2016_sachin}, dynamic bit width reduction in DCT operation \cite{tvlsi_kaushik}, or using an approximate adder \cite{TCAD_anand}. 
However, we observed that the quantization block realized using standard division algorithms consumes high power while occupying a considerable silicon area. 

For the first time, we explored an approximate quantization block by updating the Q-matrix to enable divisions with bit-shift operations, eliminating the need for high-budget standard division blocks, thereby saving energy and reducing silicon area.
Conventional approximation strategies, like loop perforation and precision scaling, are also explored in this work.
In addition to the approximate quantization block, we have proposed a heuristic-based approach to select the optimal configuration between loop perforation and precision scaling for a given quality requirement.

\subsection{Approximate Quantization}
A common approach to reducing the power of the Q block is to replace the standard division $\frac {A}{B}$ with multiplication using $A \cdot \frac{1}{B}$ using techniques like Taylor Series expansion to approximate $\frac{1}{B}$ \cite{8766885}, \cite{dac_2016_approx_divider} or reducing the width of operation in division block \cite{date_2017}. These methods, however, require one or more multipliers, which demand relatively higher energy.

By observation, quantization matrix $\mathbf{Q}$ can be replaced with approximated quantization matrix $\mathbf{Q}'$ by converting each element of the $\mathbf{Q}$ matrix to the power of 2 so that the division operation can be implemented via bit shifting. For example, $\mathbf{Q_{50}}$ can be approximated as:
\begin{equation}
    \mathbf{Q_{50}'} = 
    \label{eq:Q50_approx}
    \begin{bmatrix}
    16 & 8 & 8 & 16 & 16 & 32 & 32 & 32 \\
    8 & 8 & 8 & 16 & 16 & 32 & 32 & 32 \\
    8 & 8 & 16 & 16 & 32 & 32 & 64 & 32 \\
    8 & 16 & 16 & 16 & 32 & 64 & 64 & 32 \\
    16 & 16 & 32 & 32 & 64 & 64 & 64 & 64 \\
    16 & 32 & 32 & 32 & 64 & 64 & 64 & 64 \\
    32 & 64 & 64 & 64 & 64 & 64 & 64 & 64 \\
    64 & 64 & 64 & 64 & 64 & 64 & 64 & 64 \\
    \end{bmatrix}.
\end{equation}
Similarly, $\mathbf{Q_{90}}$ can be approximated as:
\begin{equation}
    \mathbf{Q_{90}'} = 
    \label{eq:Q90_approx}
    \begin{bmatrix}
    2 & 2 & 2 & 2 & 4 & 8 & 8 & 8 \\
    2 & 2 & 2 & 4 & 4 & 8 & 8 & 8 \\
    2 & 2 & 2 & 4 & 8 & 8 & 8 & 8 \\
    2 & 2 & 4 & 4 & 8 & 16 & 16 & 8 \\
    4 & 4 & 4 & 8 & 8 & 16 & 16 & 8 \\
    4 & 4 & 8 & 8 & 16 & 8 & 16 & 16 \\
    8 & 8 & 16 & 16 & 16 & 16 & 16 & 16 \\
    8 & 16 & 16 & 16 & 16 & 16 & 16 & 16 \\
    \end{bmatrix}.
\end{equation}

\begin{figure}[tbp]
% \centerline{\includegraphics[scale=0.35]{Metal.png}}
%\centerline{\includegraphics[width=0.48\textwidth]{distribution.png}}
\centerline{\includegraphics[width=0.45\textwidth]{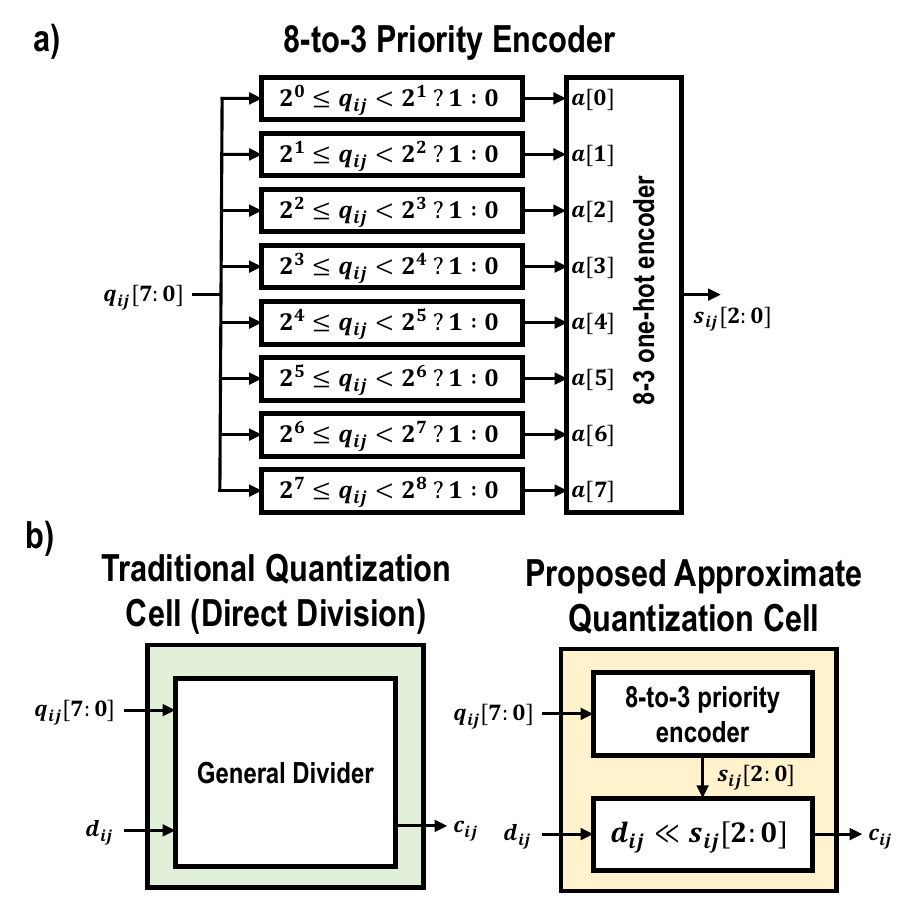}}
\caption{Hardware implementation of the proposed approximate quantization method: (a) An 8-to-3 priority encoder. (b) The proposed element-wise division cell comprises an 8-to-3 priority encoder and a barrel-shifter.}
\label{fig:QLocator}
\vspace{-0.1in}
\end{figure}

This approach introduces some errors due to approximation but reduces the divider power consumption because of the simplicity of bit shifting. Note that instead of being converted up, the elements in the original matrix are converted down to the nearest power of 2 to retain higher image quality. The general mathematical behavior of the proposed approximate technique is described as follows:
Given a quantization matrix $\mathbf{Q}$, for each element $q_{ij}$ in $\mathbf{Q}$, we first locate $q_{ij}$ using an integer $s_{ij}$ which satisfies:

\begin{equation}
    \label{eq:Qapproximation_1}
    2^{{s}_{ij}} \le {q}_{ij} < 2^{{s}_{ij}+1}
\end{equation} 
The corresponding approximated element in $\mathbf{Q'}$, $q_{ij}'$, is then constructed by:
\begin{equation}
    \label{eq:Qapproximation_2}
    {q_{ij}'} = 2^{{s}_{ij}}
\end{equation} 
If all elements in $\mathbf{Q'}$ are considered, Eq. (7) can be further extended to a matrix form:
\begin{equation}
    \mathbf{Q'} = 2 {{.}^{\wedge}} \mathbf{S}
\end{equation} 
where ${.}^{\wedge}$ is a element-wise power operator and $\mathbf{S}$ is an 8x8 matrix with its element ${s}_{ij}$ representing the exponent part of ${q}_{ij}'$. Since $1\le{q}_{ij}\le255$, we can obtain $0\le{s}_{ij}\le7$ according to Eq. (6); therefore, ${s}_{ij}$ can be represented using only 3 bits. Using this approximate technique, the quantization circuit in the JPEG compression circuit only needs to take 192 bits (3 bits x 64 elements) as the input, while the conventional division-based quantization circuit requires 512 bits (8 bits x 64 elements). At the hardware level, Eq. (\ref{eq:Qapproximation_1})(\ref{eq:Qapproximation_2}) can be implemented by an 8-to-3 priority encoder (Fig. \ref{fig:QLocator}(a)). For an 8-bit input $q_{ij}$[7:0], an 8-to-3 priority encoder can locate the first bit appearing from the MSB side and output that particular location by a 3-bit signal  $s_{ij}$[2:0]. The quantization is then performed by a bit shifter, with $s_{ij}$[2:0] being the shifting amount. Therefore, our proposed architecture realizes the approximated element-wise division operation through an 8-to-3 priority encoder and a barrel-shifter, as shown in Fig. \ref{fig:QLocator}(b).
Note that at the time of decoding, the same quantization matrix $\mathbf{Q'}$ must be used for better reconstruction of the image.  

\subsection{Precision Scaling}
Precision scaling, or bit truncation, alleviates the computational load of image compression by reducing the data width of the input image. LSB (Least Significant Bit) truncation is realized in this paper. The size of data reduction, \textit{truncation level} $B_j$, has to be specified before image compression begins. The truncated pixel block, $\mathbf{{M}_{tr}}$, can be described in terms of the original pixel block $\mathbf{M}$ and truncation level $B_j$ as
\begin{equation}
    \mathbf{{M}_{tr}} = round\{\frac{1}{{2}^{B_j}}\mathbf{M}\}
\end{equation}
The precision scaling technique allows the DCT and quantization to function with fewer bits overhead. \textcolor{black}{Note that bit truncation is implemented uniformly throughout the data path of the hardware accelerator. Each data-path component (adders, multipliers, etc.) can be equipped to modify the operational bit-width by introducing a bit-wise clock gating in each of them, thereby reducing energy consumption.} %\textcolor{red}{FIGURE (Num)} shows the case of both LSB truncation and MSB truncation, with the entire data width $n=8$ and the truncation level $k=1$. 

\subsection{Loop Perforation}
\begin{comment}
    To exploit spatial redundancies in an image, loop skipping (or loop perforation) is employed to skip the compressing process for the current pixel block if it is similar to the previous one. This significantly reduces the energy dissipated for computation while maintaining acceptable degradation in the encoded image quality.
\end{comment}

Loop perforation, or loop skipping, takes advantage of spatial redundancies in an image. This technique involves bypassing the compression process for the current pixel block if it closely resembles the preceding one. This results in a substantial decrease in the energy expended for computation while preserving an acceptable level of degradation in the encoded image quality. \par
This work implements the loop skipping function by first asserting the loop skip threshold $L_{i} = i$ ($i$ is a non-negative integer), representing the error tolerance $\varepsilon$ with  $\varepsilon$ = $5i$. To determine whether a pixel block should be sent into the JPEG compression computation core, the pixel block is compared with the previously processed pixel block by checking if all pixels in two blocks are within the error tolerance. The exact step-by-step operation is illustrated in Algorithm \ref{alg:lookskip}. If a pixel block satisfies the loop skipping criteria, the JPEG compression circuit will disable the computation core and directly output the computation result of the previously processed pixel block. \par
\begin{figure}[tbp]
% \centerline{\includegraphics[scale=0.35]{Metal.png}}
%\centerline{\includegraphics[width=0.48\textwidth]{distribution.png}}
\centerline{\includegraphics[width=0.5\textwidth]{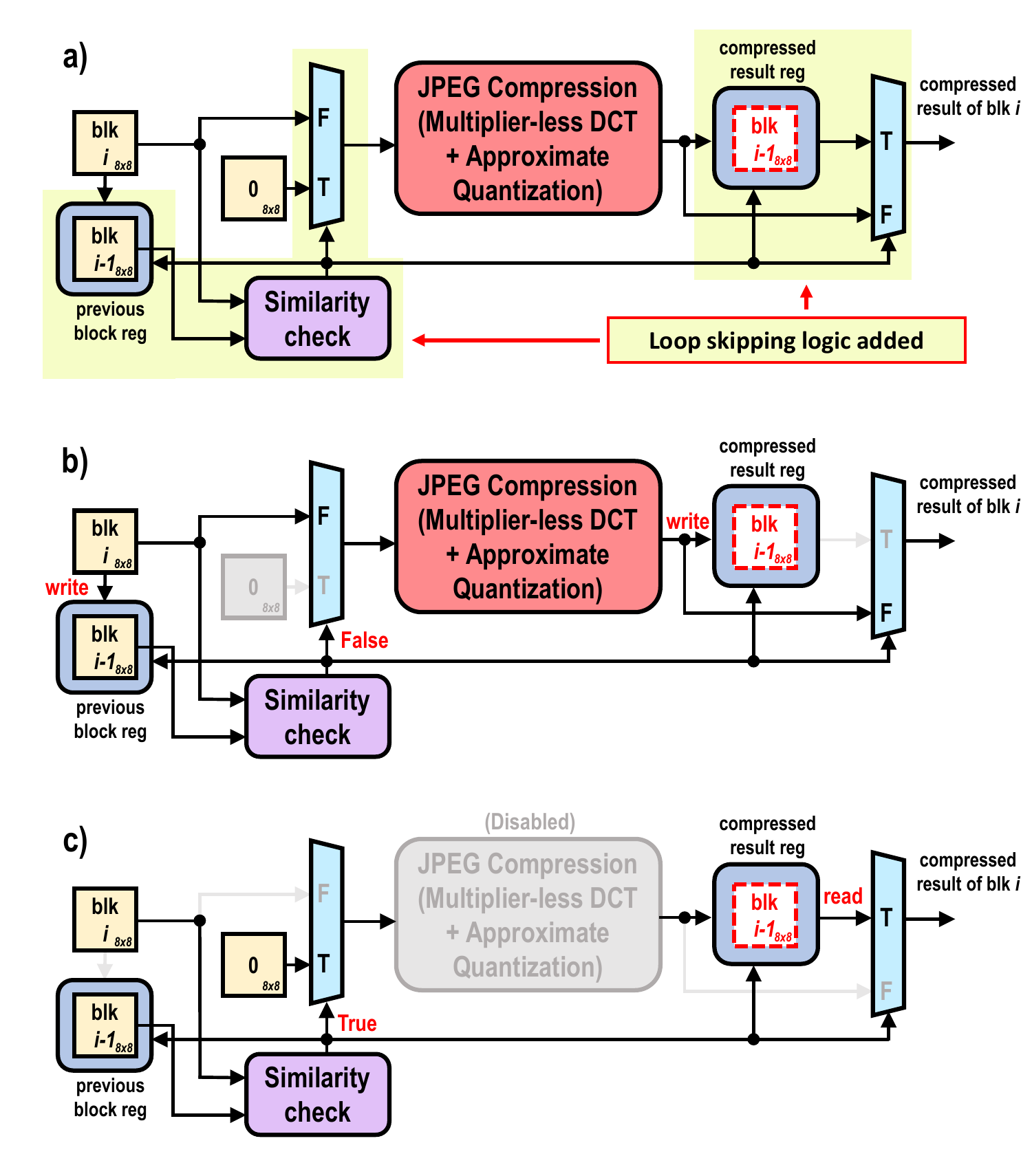}}
\caption{Hardware realization of loop perforation: a) additional registers and logic added, b) data flow when the similarity of neighboring blocks is not detected, and c) data flow when the similarity of neighboring blocks is detected. }
\label{fig:LoopSkipBlockDiagram}
\vspace{-0.1in}
\end{figure}

Fig. \ref{fig:LoopSkipBlockDiagram}(a) shows the additional hardware cost incurred to perform loop skipping, where registers for storing the previous block and its corresponding results, a similarity checker, and related selection logic (MUX) are added to the original JPEG core. Fig. \ref{fig:LoopSkipBlockDiagram}(b)(c) provide insight into the data flow and circuit operation under conditions where the similarity of two blocks is detected or not. Fig. \ref{fig:LoopSkipBlockDiagram}(b) depicts the situation where no similarity is detected. If two neighboring blocks are not similar enough, the similarity check logic will generate a FALSE signal, writing 64 pixels of the current block into the previous block register and saving the current block compressed results for the next cycle comparison. On the other hand, if the similarity is detected, a TRUE signal will be sent out from the similarity check logic, which disables the JPEG core and outputs the computation results directly from the compressed result registers. This scenario is illustrated in Fig. \ref{fig:LoopSkipBlockDiagram}(c). \par 
\begin{comment}
  Although to perform loop perforation on the circuit level, additional hardware resources, such as the logic that decides if two-pixel blocks are similar and registers that store the previously-processed pixel block and corresponding results, have to be added, the overall power saving is still significant because the DCT unit, a much more power-consuming block, is disabled.   
\end{comment}

The overall power savings remain substantial despite the need for additional hardware resources at the circuit level to implement loop perforation, such as the logic determining the similarity of the pixel blocks and registers storing the previously processed pixel block and its corresponding result. This is primarily due to the disabling of the DCT unit, a significantly more power-consuming block.
This point will be further discussed in Section \ref{sec:result_all}. \par By increasing the loop skipping levels, higher energy savings are achieved at the expense of degraded image quality. In this work, we have implemented loop skipping in the software and the hardware that performs the desired operations and generates the required control signals before the acceleration.

%% ===============================================

%% ===============================================

\LinesNumbered
%\IncMargin{1.5em}
\normalem
\begin{algorithm}[!th]
%\footnotesize
	%\SetAlgoLined
	\caption{\textbf{Loop skipping logic}}
	\label{alg:lookskip}
	%\Indm
	\KwIn{the on-trial 8x8 block $\mathbf{M_{in}}$, the latest-processed block $\mathbf{M_{l}}$ and its JPEG compression result $\mathbf{C_{l}}$, and the error tolerance $\varepsilon$}
	\KwOut{the new latest-processed block $\mathbf{M_{out}}$ and its corresponding result $\mathbf{C_{out}}$}
	%Initialize: isBlockSimilar = 1;\\ 
    \For{($i=0; i<8; i=i+1$)}{
 	\For{($j=0; j<8; j=j+1$)}{
            %\While{isBlockSimilar} {
    		  ceiling $ = \min \{ {m_{l,ij}} + \varepsilon, 127 \} $;  \\
                floor $ = \max \{ {m_{l,ij}} - \varepsilon, -128 \} $;  \\
                \If{ (($m_{in,ij}$ $>$ ceiling) or ($m_{in,ij}$ $<$ floor)) } {
                    %isBlockSimilar = 0;\\
                    $\mathbf{M_{out}}$ = $\mathbf{M_{in}}$; \\
                    $\mathbf{C_{out}}$ =  $JpegCompression$($\mathbf{M_{in}}$);\\
                    \Return $\mathbf{M_{out}}$, $\mathbf{C_{out}};$
                }        
            %}
        }
    }
     $\mathbf{M_{out}}$ = $\mathbf{M_{l}}$; $\mathbf{C_{out}}$ = $\mathbf{C_{l}}$; \\
	\Return $\mathbf{M_{out}}$, $\mathbf{C_{out}};$

\scriptsize
\end{algorithm}
\ULforem

\subsection{Dynamic selection of optimum approximation technique}

To maximize the energy savings from approximate computing, we propose to combine the \textit{loop skipping} and \textit{precision scaling} techniques. The combination of both strategies performs better than either standalone scheme.

\LinesNumbered
%\IncMargin{1.5em}
\normalem
\begin{algorithm}[!th]
%\footnotesize
	%\SetAlgoLined
	\caption{\textbf{To extract the \emph{Q-E} characteristics for individual approximation technique}}
	\label{alg:qe_single}
	%\Indm
	\KwIn{Set of required image qualities: $Q[0:N-1]$} 
	\KwOut{Set of quality knob configuration: $k[0:N-1]$ and energy consumption: $E[0:N-1]$ corresponding to $Q[0:N-1]$}
	\For{($i=0;i<N;i=i+1$)}{ 
		m = 0; \\
		\While{($(k[m]) \ge Q[i]$)}{ 
			m = m+1;%\tcp*[r]{increase degree of approx until Q violation}
		}
		$E[i]=E[(k[m-1])]$; 
            $k[i] = k[m-1]$;}
	\Return $k, Q, E$

\scriptsize
\end{algorithm}
\ULforem

\begin{figure}[t]
\centering
\includegraphics[width=\columnwidth]{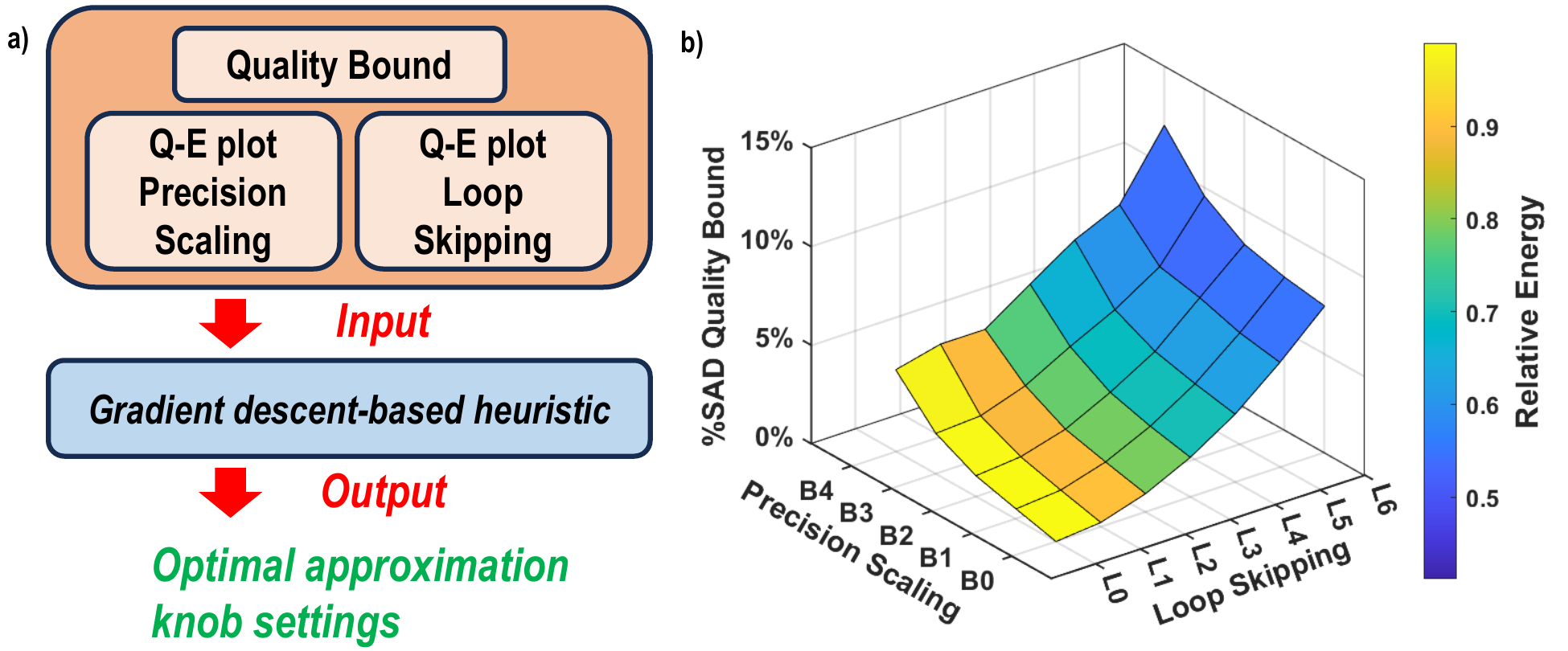}
\caption{\label{fig: heuristic} Gradient Descent Algorithm: (a) block diagram and (b) 3D Quality vs. Energy plot. }
\vspace{-0.1in}
\end{figure}

A gradient descent-based heuristic algorithm determines the optimal approximation degrees of bit truncation and loop skipping using the Quality-Energy (Q-E) plots of individual approximation strategies, as suggested by Fig.~\ref{fig: heuristic}(a). To quantify the effect of approximation on the quality of the image, SAD (Sum of Absolute Differences) is chosen as a performance metric, which is defined as the ratio of the sum of absolute differences in pixel values between the generated and the reference image to the sum of pixel values in the reference image. Note that \%SAD degradation has shown a good correlation to other metrics such as PSNR (Peak Signal to Noise Ratio) and SSIM (Structural Similarity) and is used for its easy computation in the heuristics and lower overhead.  Also, note that the quality of the image is inversely proportional to this metric. However, we use SSIM and PSNR while evaluating the approximation techniques, as those metrics are well-accepted in the image processing community.

\begin{figure}[t]
\centering
\includegraphics[width=1.1\columnwidth]{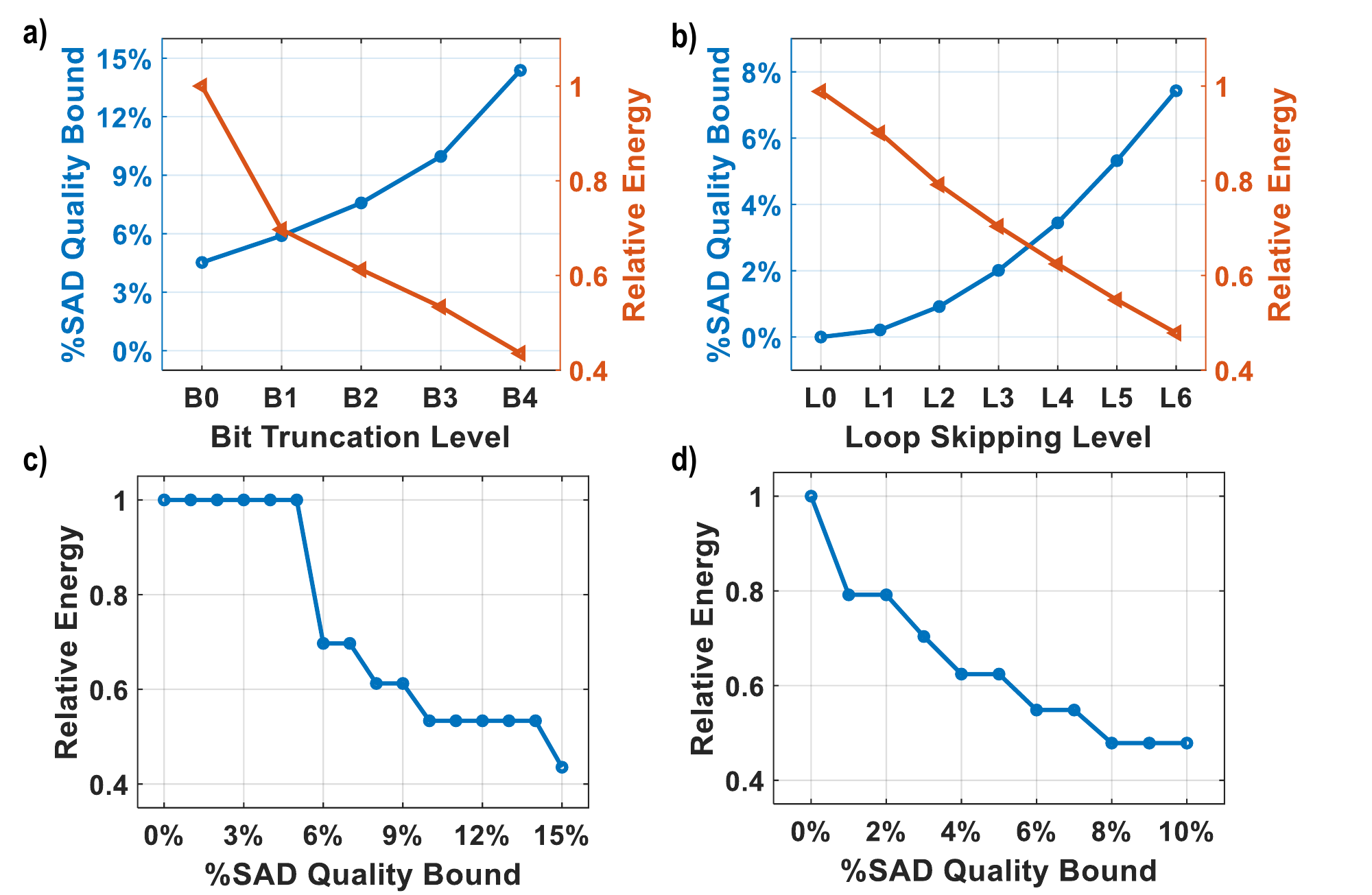}
\caption{\label{fig:LS_BT} (a)-(b): Normalized energy consumption vs SAD degradation bound for (a) Bit Truncation and (b) Loop skipping. (c)-(d): Individual E plots for (c) Bit Truncation and (d) Loop Skipping generated using Algorithm \ref{alg:qe_single}. }
\vspace{-0.1in}
\end{figure}

Algorithm~\ref{alg:qe_single} obtains the individual Q-E plots that provide insights into the degradation of the quality of the decoded images for benefits in relative energy for different approximation scenarios. For a particular output image quality bound, the quality knobs $B_0-B_4$ and $L_0-L_6$, along with the relative energy savings, are found by the online image gallery of the Computer Vision Group of the University of Granada \cite{IMGdataset}.
Fig.~\ref{fig:LS_BT} shows the extracted plots for loop perforation and precision scaling.

Algorithm~\ref{alg:qe} employs a gradient descent-based optimization search using these extracted plots to provide an overall Quality vs. Energy for the combined strategy. We vary the loop perforation categories ($L_i$) and bit truncation levels ($B_j$) to obtain the optimum settings for a particular output quality bound ($Q_{A}$). In other words, we are searching for the optimal solution ($\hat B_{i}, \hat L_{i}$) such that
\begin{equation}
%\begin{aligned}
    \begin{tabular}{rl} \label{eq:GDA}
        minimize & $E(B_{i},L_{i})$\\
        subject to & $Q(B_{i},L_{i}) \leq Q_{A}$\\
                   & $Q(B_{i},L_{i}) \geq 0$\\
                   & $B_{i} \geq 0, \; L_{i} \geq 0$
    \end{tabular}
\end{equation}

where $E(B_{i},L_{i})$ and $Q(B_{i},L_{i})$ represent the relative energy and \%SAD degradation under particular $B_{i}$ and $L_{i}$, respectively. The convexity of the problem can be justified as follows: First, $Q(B_{i},L_{i})$ is monotonically increasing in each dimension because higher bit truncation or loop skipping levels result in higher \%SAD degradation. Second, since the relative energy and quality degradation are two inversely related variables, to minimize $E(B_{i},L_{i})$ is equivalent to maximizing $Q(B_{i},L_{i})$. With these two properties, Eq. (\ref{eq:GDA}) can be treated as a maximization problem constrained in the first octant of the 3-D space expanded by $B_{i}$, $L_{i}$, and $Q(B_{i},L_{i})$, where the objective function is monotonically increasing in the direction of $B_{i}$ and $L_{i}$. As a result, the convexity is assured, and it is feasible to apply a gradient descent algorithm to find the optimal.
 
The controller implementing this heuristic, realized in software code, automatically configures the degree of loop perforation and bit truncation, $L_i$ and $B_j$, respectively, by moving in the direction of the steepest gradient of the ratio of energy savings to quality degradation resulting from the variation in each degree of the approximation knobs. 
\textcolor{black}{Fig.~\ref{fig: heuristic}(b) shows the 3D Q-E plot for different $L_i$ and $B_j$, along with the relative energy required for processing. For a specified quality degradation bound, our proposed gradient descent algorithm uses the 3D plot and selects the best ($B_i$, $L_j$) combination that results in the lowest energy configuration according to the color bar on the right according to Algorithm~\ref{alg:qe}.}

\LinesNumbered
%\IncMargin{1.5em}
\normalem
\begin{algorithm}[!th]
%\footnotesize
	%\SetAlgoLined
	\caption{\textbf{Gradient descent determines the optimal approximation degrees for a given quality bound}}
	\label{alg:qe}
	%\Indm

\KwIn{Output quality bound: $Q_A$, \\ Quality \emph{vs.} Energy (\emph{Q-E}) curves for loop perforation and precision scaling $(Q_{l}$-$E_{l})$ and  $(Q_{t}$-$E_{t})$, respectively}
	\KwOut{Optimal approximation knob settings ($i$, $j$) according to $Q_A$}
	\Indp
	Initialize: $i=j=0$, $Q=1$, $E=1$\\ %\tcp*[r]{estimate initial x from char-plot}
	\While{($Q\ge Q_A$)}{$E_{l\Delta}$=$E_{l}[i]$-$E_{l}[i+1]$; $Q_{l\Delta}$=$Q$-$Q_{l}[i+1]$;\\
	$E_{t\Delta}$=$E_{t}[j]$-$E_{t}[j+1]$; $Q_{t\Delta}$=$Q$-$Q_{t}[j+1]$;\\
	\If{($\frac{Q_{l\Delta}}{E_{l\Delta}} \ge \frac{Q_{t\Delta}}{E_{t\Delta}}$)}{
    \If {($Q_t[j+1]$ $\le$ $Q_A$)} 
        {$E=E-E_{t\Delta}$;
        $Q=Q-Q_{t\Delta}$;
        $j=j+1$;}
    \ElseIf {($Q_l[i+1]$ $\le$ $Q_A$)}{ 
    $E=E-E_{l\Delta}$;
		 $Q=Q-Q_{l\Delta}$; $i=i+1$;}
	}
%	\\\If {($Q$ $\le$ $Q_A$)} {$E=E+E_{l\Delta}$;
%		\\$Q=Q+Q_{l\Delta}$; \\ \textbf{break;}}
	%\\{$E=E+E_{t\Delta}$;
	%	\\$Q=Q+Q_{t\Delta}$; \\ \textbf{break;}}
    \Else {
    \If {($Q_l[i+1]$ $\le$ $Q_A$)}{ 
    $E=E-E_{l\Delta}$; 
    $Q=Q-Q_{l\Delta}$; 
    $i=i+1$;}
%	\\\If {($Q$ $\le$ $Q_A$)} {$E=E+E_{l\Delta}$;
%		\\$Q=Q+Q_{l\Delta}$; \\ \textbf{break;}}
    \ElseIf {($Q_t[j+1]$ $\le$ $Q_A$)} 
      {$E=E-E_{t\Delta}$;
        $Q=Q-Q_{t\Delta}$;
        $j=j+1$;}
    }
    }
    \Return $i$, $j$;
\scriptsize
\end{algorithm}
\ULforem
\vspace{0.1in}

%%KGK: Give heuristic, we will estimate hardware cost for this too. A nice 3-D plot will be helpful.
\begin{figure}[htbp]
\centerline{\includegraphics[width=0.45\textwidth]{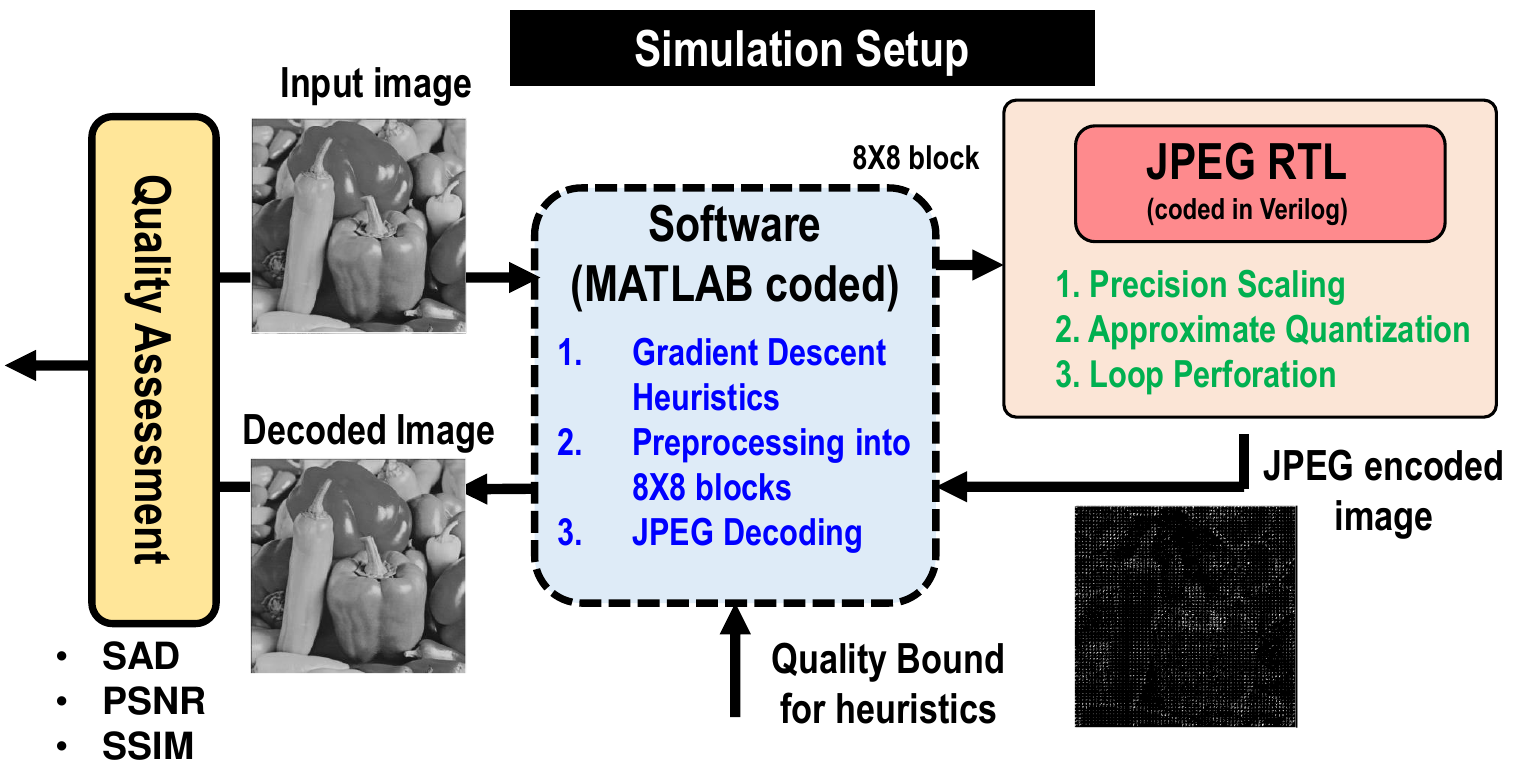}}
\caption{Simulation setup.}
\label{sim_setup}
\vspace{-0.2in}
\end{figure}

\section{Simulation Methodology}\label{sec:sim_setup}
Fig.~\ref{sim_setup} depicts the simulation setup used for testing the functionality of the hardware. Our simulation is conducted over an image dataset from the gallery of the Computer Vision Group of the University of Granada \cite{IMGdataset}, where many representative images in the field of image processing, such as Baboon, Boat, Barbara, Pirate, Bridge, and Airplane, are included. The input image is reduced to chunks of 8$\times$8 matrix in MATLAB before feeding to the design under test (DUT). The MATLAB code runs the gradient descent-based heuristic algorithm to estimate the optimal approximation knobs for a particular input quality bound (SAD). %The loop perforation is implemented in software to compare the current set of pixels with the previous set to skip the computation based on the spatial locality of the pixels ($L_i$).% 
The software also decides the degree of precision scaling to be realized and configures the hardware by clock gating the required bits throughout the accelerator. 
A Verilog test bench is used to convey the appropriate degree of truncation and the pre-processed image (in a text file) for simulation.
The accelerator performs the JPEG encoding on the input image and writes the output processed image in a separate text file, which is then reconstructed through inverse quantization and inverse DCT in the software. Inverse quantization is an operation of element-wise multiplication, which is given by:
\begin{equation}
\mathbf{R} = \mathbf{C} \odot \mathbf{Q}
\end{equation}
where $\odot$ is the element-wise multiplication operator, and $\mathbf{R}$ is the result of inverse quantization. \par In general, the $\mathbf{Q}$ used in this step should be the same one as used in the encoding process. However, to give a more comprehensive analysis of the effect of introducing an approximated quantization matrix, we also analyze the case where the approximated matrix $\mathbf{Q}'$ is used in encoding while the unmodified one $\mathbf{Q}$ is still used for inverse quantization. \par
Inverse DCT is given by the transformation of:
\begin{equation}
\mathbf{N} = \mathbf{T^{'}RT}
\end{equation}
where $\mathbf{N}$ is the result of inverse DCT and $\mathbf{T}$ is given by Eq. (\ref{eq:DCT_matrix}). The reconstructed image can be obtained after rounding $\mathbf{N}$'s all entries. In the end, performance evaluation and quality assessment are conducted based on the reconstructed results.
\par To validate the accelerator's functionality, we use in-built MATLAB-based JPEG compression and compare it with hardware output.
The JPEG RTL is synthesized using Synopsys Design Compiler, mapped to TSMC 65nm standard cell library. The functionality of the extracted netlist is re-validated. All the RTL simulations are performed using the Cadence NC-Verilog simulator. 
The design area and power values are provided from the post-synthesis simulation results. Note that results from Spice simulations (done in Cadence Virtuoso) are utilized as they provide precise energy numbers. 

\section{Results}\label{sec:sim_results}
In this section, we first discuss the effect of individual approximation techniques on the overall performance of JPEG compression hardware. The performance of the combined approximation strategy that dynamically tunes the configuration of the constituent techniques is also shown. Note that system-level hardware performances like power and area are not reported in previous related literature, such as \cite{Multi-Object_Application-Driven_Approx_Design_Method} (which applies bit-truncation and loop peroration in JPEG compression) or \cite{barbareschi2022genetic} (which works on the approximation of DCT hardware). However, this work provides the power and area numbers for the most optimal design, including the DCT, approximate quantization block, and loop-skipping circuitry from the post-synthesis Spice simulations. The quality of the images is evaluated in terms of SSIM, PSNR, and SAD as discussed earlier in Section ~\ref{sec:approx_tech}.

\subsection{Approximate Quantization}

\begingroup
\begin{table*}[h]
  \centering
\caption {\label{tab:approx_Q_ssimpsnr}  Statistics of Fig. \ref{fig:approx_Q_ssimpsnr}: Image quality under different quantization schemes.} 
%\begin{ruledtabular}
\begin{tabular}{cccccccccccc}\hline 
\multirow{2}{*}{Quality Level} &  & \multicolumn{4}{c}{SSIM} & & \multicolumn{4}{c}{PSNR}  
        \\ \cline{3-6} \cline{8-11}  
        & & mean & std & max & min & & mean & std & max & min \\  \hline
\multirow{3}{*}{Q50}  & Division   & $0.8944$  & $0.0327$  & $0.9472$ & $0.7811$  & & $32.17$  & $2.99$ & $37.20$ & $24.51$   \\       
& Bit shifting   & $0.9137$  & $0.0252$  & $0.9578$ & $0.8387$   & & $33.30$  & $2.80$ & $38.10$ & $26.49$   \\
 & $\%$ Increase  & $2.19\;\%$  & $1.22\;\%$  & $7.37\;\%$ & $0.99\;\%$   & & $3.62\;\%$  & $1.24\;\%$ & $8.09\;\%$ & $2.07\;\%$   \\ \hline
\multirow{3}{*}{Q90} & Division   & $0.9689$  & $0.0092$  & $0.9934$ & $0.9429$   & & $38.98$  & $1.62$ & $42.17$ & $36.56$   \\       
 & Bit shifting   & $0.9770$  & $0.0069$  & $0.9961$ & $0.9574$   & & $40.57$  & $1.35$ & $43.31$ & $38.77$   \\
 & $\%$ Increase  & $0.84\;\%$  & $0.3\;\%$  & $1.56\;\%$ & $0.26\;\%$   & & $4.11\;\%$  & $0.94\;\%$ & $6.14\;\%$ & $2.63\;\%$   \\ \hline
\end{tabular}
%\end{ruledtabular}
\end{table*}
\endgroup

\subsubsection{Case I: images reconstructed by the corresponding encoding quantization matrix (Eq.~(\ref{eq:Q50_approx})(\ref{eq:Q90_approx}))}

\begin{figure}[t]
\centerline{\includegraphics[width=0.5\textwidth]{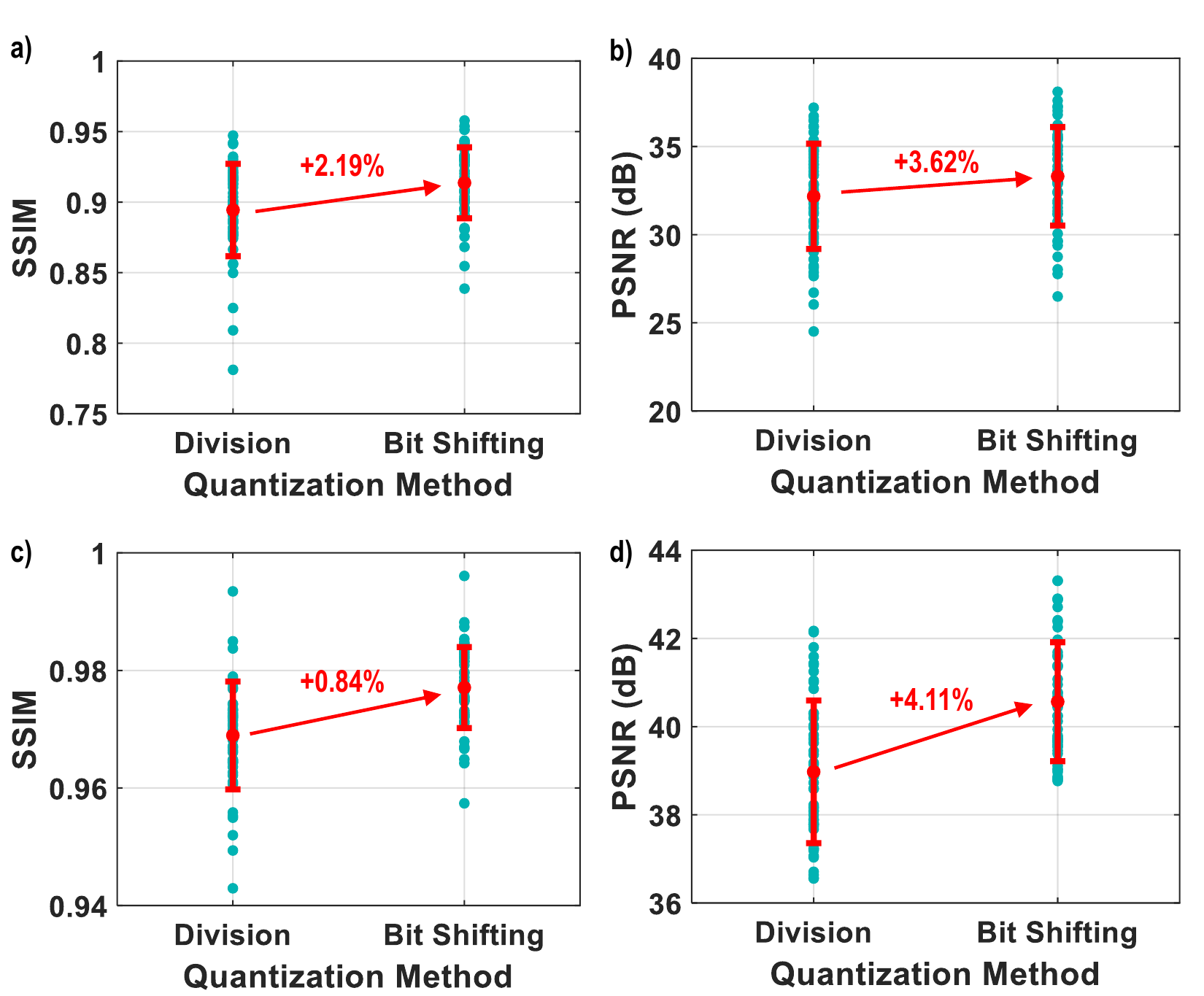}}
\caption{Effect of quantization block approximation on image quality: (a)-(b) SSIM and PSNR comparison for Q50, and (c)-(d) SSIM and PSNR comparison for Q90.}
\label{fig:approx_Q_ssimpsnr}
%\vspace{-0.2in}
\end{figure}

The scatter plots from Fig.~\ref{fig:approx_Q_ssimpsnr} and the corresponding statistic value in Table~\ref{tab:approx_Q_ssimpsnr} together show the change in the SSIM and PSNR of the reconstructed images because of the use of an approximation matrix. It is observed that the reconstructed images encoded using bit-shifting-based quantization demonstrate better quality (higher SSIM and higher PSNR) than the reconstructed images encoded using standard division-based quantization for quality levels 50 and 90. This is because every element in the quantization matrix is down-approximated to its closest power of 2, resulting in a new quantization matrix whose quality factor is higher than the original's. This result, however, entails a reduction in the compression ratio of the compressed image. \par

\begin{figure}[tbp]
\centerline{\includegraphics[width=0.5\textwidth]{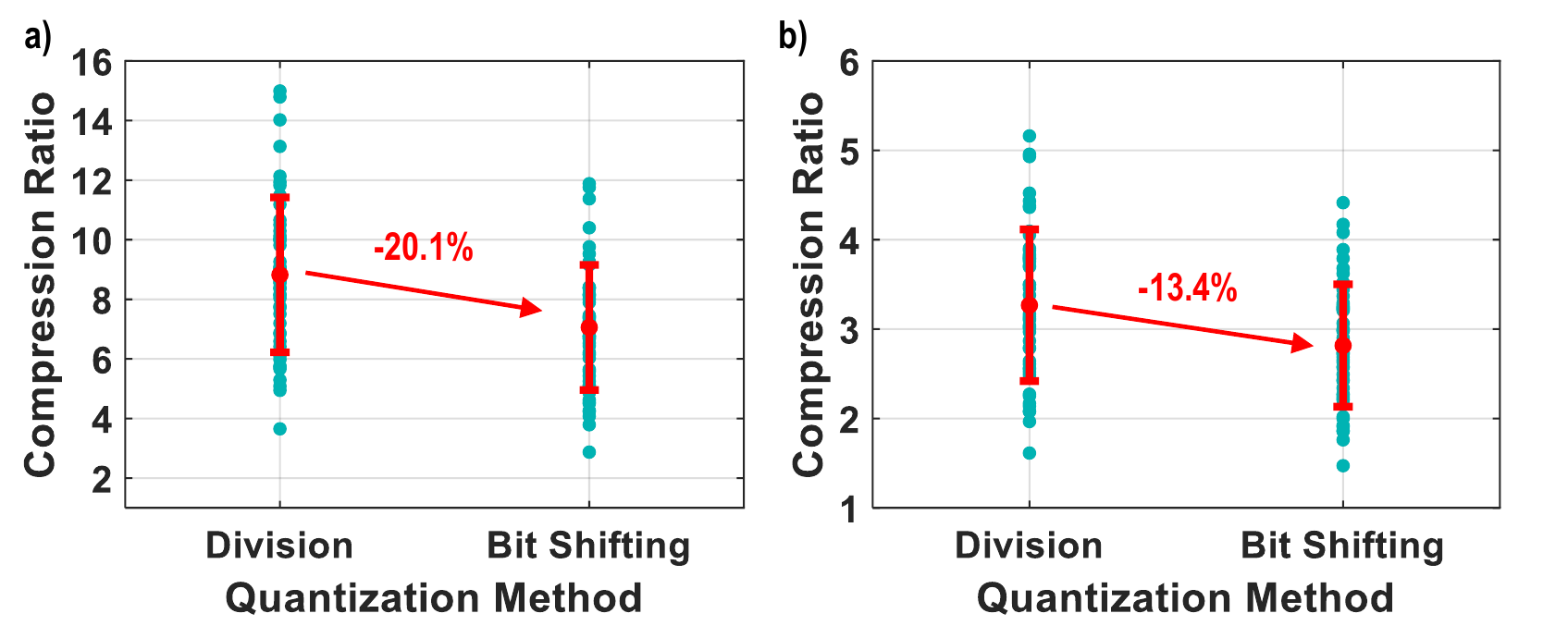}}
\caption{Effect of quantization block approximation on compression ratio: (a) on Q50 and (b) on Q90.}
\label{fig:approx_Q_compressionratio}
%\vspace{-0.15in}
\end{figure}

\begin{figure}[htbp]
\centerline{\includegraphics[width=0.32\textwidth]{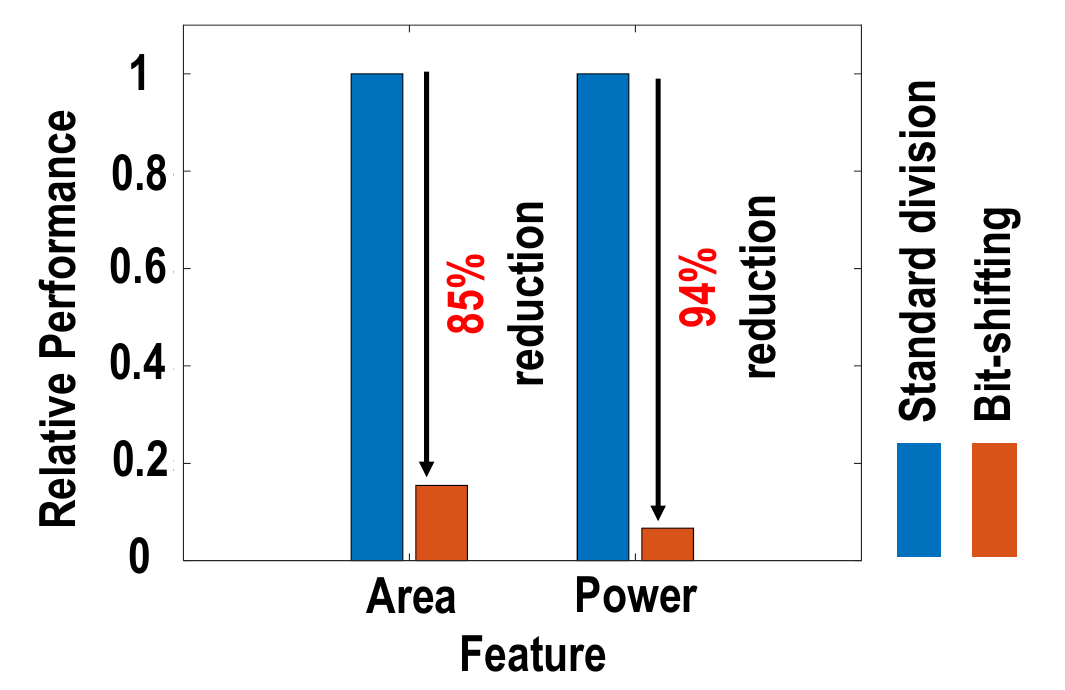}}
\caption{ Hardware-level area and power improvement based on approximate quantization: bit-shifting-based quantization vs. division-based quantization.}
\label{fig:AreaPower_Q_plot}
%\vspace{-0.1in}
\end{figure}

Fig.~\ref{fig:approx_Q_compressionratio} shows the effect of the approximated quantization method on compression ratio, where the $20.1\%$ and $13.4\%$ compression ratio decline are observed when approximated Q50 and Q90 are adopted, respectively. Detailed statistic values of Fig.~\ref{fig:approx_Q_compressionratio} are provided in Table~\ref{tab:approx_Q_compressionratio}. Although there are reductions in compression ratio, the advantages of using approximated quantization matrices are evident once the hardware-level implementation is considered. Fig.~\ref{fig:AreaPower_Q_plot} enlightens the benefits in power and area using the proposed quantization scheme, achieving $85\%$ reduction in area and $94\%$ power savings for conventional division-based quantization block.

\begingroup
\begin{table}[h]
  \centering
\caption {\label{tab:approx_Q_compressionratio} Statistics of Fig. \ref{fig:approx_Q_compressionratio}: compression ratio under different quantization schemes.} 
%\begin{ruledtabular}
\begin{tabular}{cccccc}\hline 
\multirow{2}{*}{Q Matrix }  & & \multicolumn{4}{c}{Compression Ratio}
        \\ \cline{3-6}   
        & & mean & std & max & min  \\  \hline
\multirow{3}{*}{Q50} & Division   & $8.82$  & $2.6$  & $14.99$ & $3.65$    \\       
 & Bit shifting   & $7.06$  & $2.1$  & $11.88$ & $2.87$      \\
 & Degradation  & $20.1\;\%$  & $1.85\;\%$  & $17.13\;\%$ & $26.4\;\%$  \\ \hline
\multirow{3}{*}{Q90} & Division   & $3.27$  & $0.89$  & $5.16$ & $1.61$       \\       
 & Bit shifting   & $2.82$  & $0.68$  & $4.41$ & $1.47$      \\
 & Degradation  & $13.4\;\%$  & $1.94\;\%$  & $8.77\;\%$ & $17.34\;\%$      \\ \hline
\end{tabular}
%\end{ruledtabular}
\end{table}
\endgroup
 
\subsubsection{ Case II: images reconstructed by the unmodified quantization matrix (Eq.~(\ref{eq:Q50_unmodified})(\ref{eq:Q90_unmodified}))}
\begin{figure}[tbp]
\centerline{\includegraphics[width=0.51\textwidth]{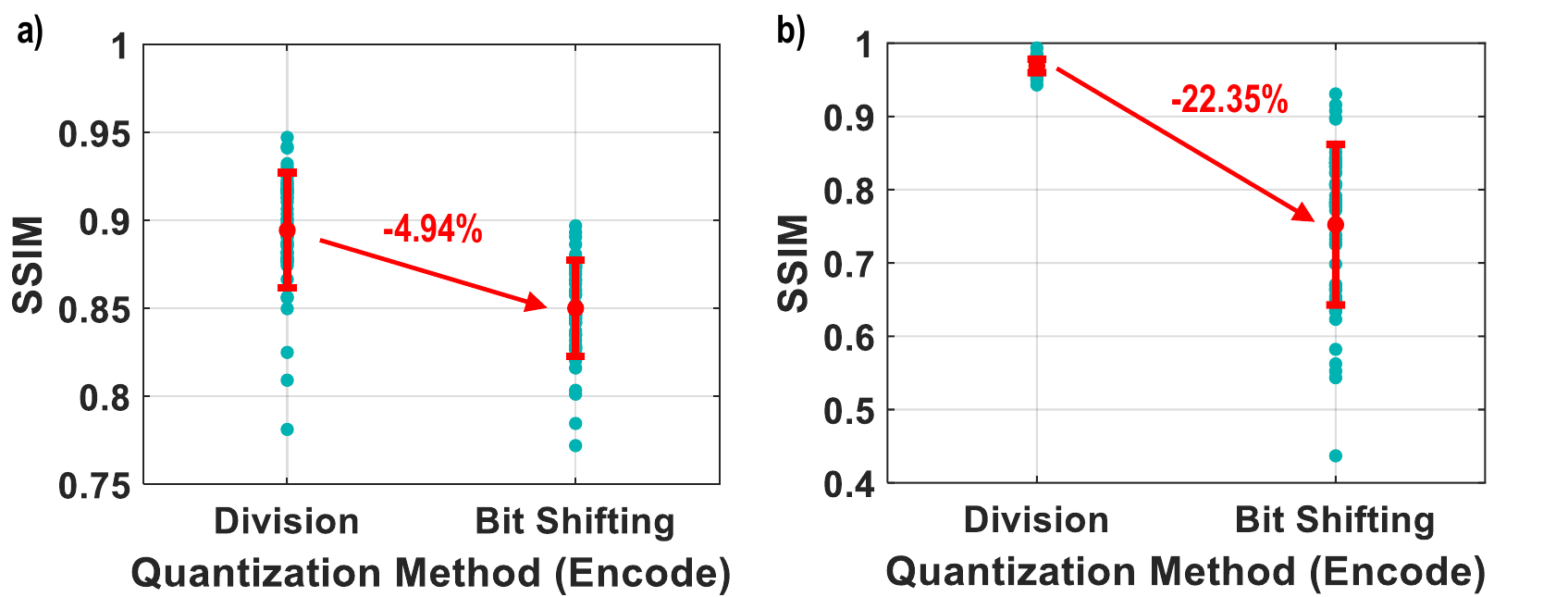}}
\caption{Comparison of reconstructed image quality between the image decoded by the standard Q and by the modified Q for (a) Q50 and (b) Q90.}
\label{fig:Qcompare_EnQapDeQap_vs_EnQapDeQstd}
\vspace{-0.1in}
\end{figure}

One should decode the compressed image with the same quantization matrix used in the encoding stage for better image quality when reconstructing an image. However, since the approximate quantization circuit is only applied in the encoding end in this work, we assume that the decoding end may still use the unmodified quantization matrix to reconstruct images. The discussion about using different quantization matrices for encoding and decoding is thus presented in Fig.~\ref{fig:Qcompare_EnQapDeQap_vs_EnQapDeQstd}, which provides the image quality comparison between images reconstructed with the modified (approximated) $\mathbf{Q}$ matrix and with the original (standard) one for $\mathbf{Q_{50}}$ and $\mathbf{Q_{90}}$. Using the standard matrix $\mathbf{Q_{50}}$ to reconstruct the image encoded by the approximated matrix $\mathbf{Q_{50}'}$ results in $4.94\%$ SSIM degradation, as shown in Fig.~\ref{fig:Qcompare_EnQapDeQap_vs_EnQapDeQstd}(a). However, in the case of $\mathbf{Q_{90}}$, using the standard $\mathbf{Q_{90}}$ to reconstruct images induces more SSIM degradation than using the standard $\mathbf{Q_{50}}$, as shown in Fig.~\ref{fig:Qcompare_EnQapDeQap_vs_EnQapDeQstd}(b), where $22.35\%$ SSIM degradation is presented.\par

This result stems from the fact that the first entry in Q50 is originally a 2's power ($\mathbf{Q_{50}}$(0,0) = 16), as is not the case for Q90 ($\mathbf{Q_{90}}$(0,0) = 3). For the scenario where $\mathbf{Q_{90}'}$ is used in encoding while $\mathbf{Q_{90}}$ is used in decoding, different divisors for the first entry of the DCT coefficient block (also known as the DC coefficient) are used in the quantization ($\mathbf{Q_{90}'}$(0,0) = 2), and inverse quantization ($\mathbf{Q_{90}}$(0,0) = 3). This discrepancy corrupts the reconstructed value of the DC coefficient at the step of inverse quantization. Since for still images, most of the energy is located in the low-frequency area \cite{DCTenergyexplanation}, the corrupted DC coefficient will then result in severe image degradation after inverse DCT is performed. Note that this problem could be addressed by keeping the first entry in the standard $\mathbf{Q_{90}}$ matrix unmodified and designing a multiplier-less divider. For example, the element-wise division for $\mathbf{Q_{90}}$'s first entry (quantizing a number by 3) can be approximately implemented as 1/4+1/8-1/16+1/32 (0.334). Thus, this can be implemented just by addition and subtraction along with bit-shift operation.

\begin{figure}[tbp]
\centerline{\includegraphics[width=0.4\textwidth]{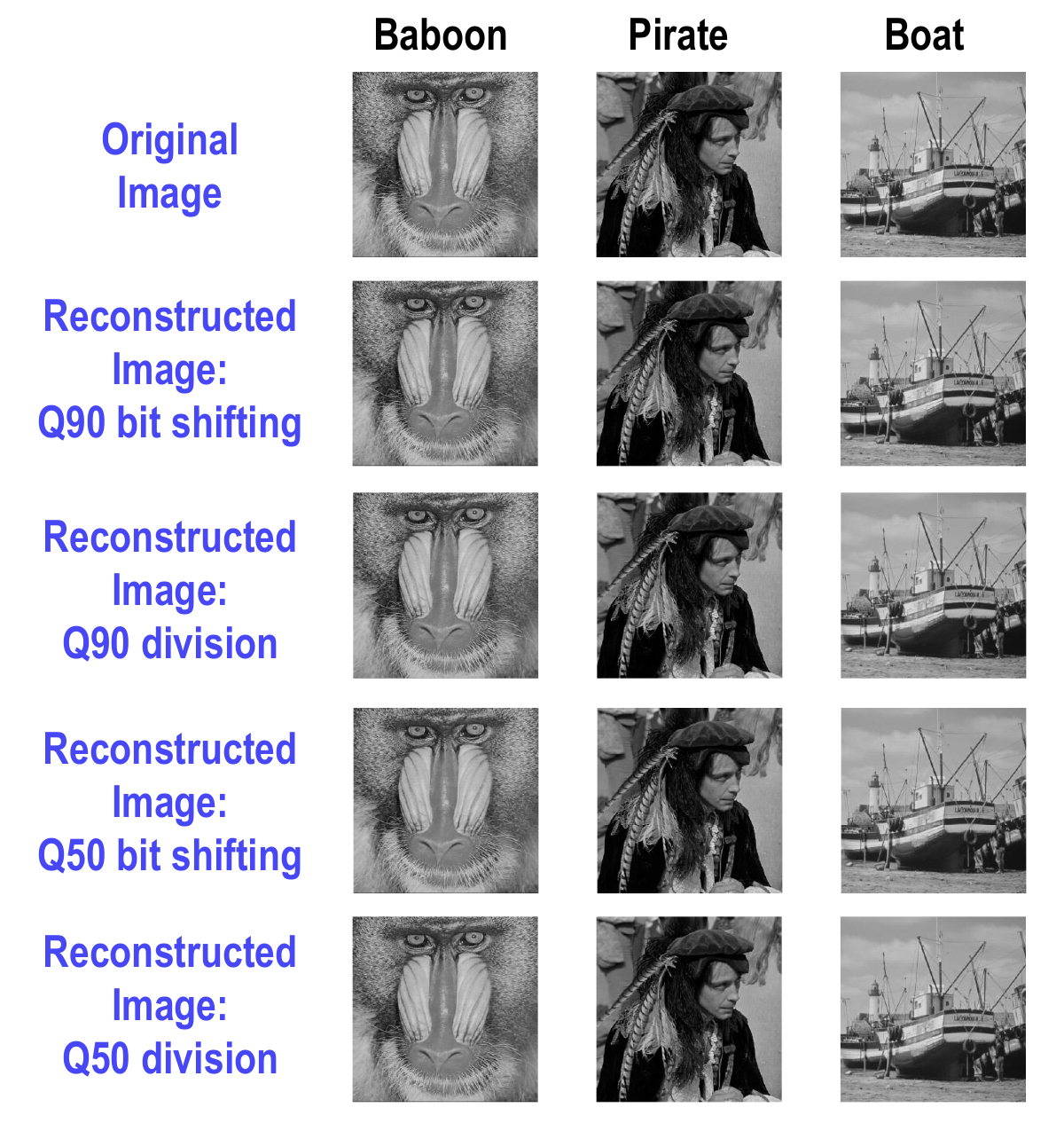}}
\caption{Reconstructed images using approximate division block vs. standard division block with two different quality levels 50 and 90.}
\label{fig:approx_Q_results}
%\vspace{-0.1in}
\end{figure}
Lastly, we present subjective analysis on three images (Baboon, Pirate, and Boat) selected from the image dataset. Fig. ~\ref{fig:approx_Q_results} shows the reconstructed images using different quantization methods and levels.

\subsection{Precision scaling}
\begin{figure}[htbp]
\centerline{\includegraphics[width=0.52\textwidth]{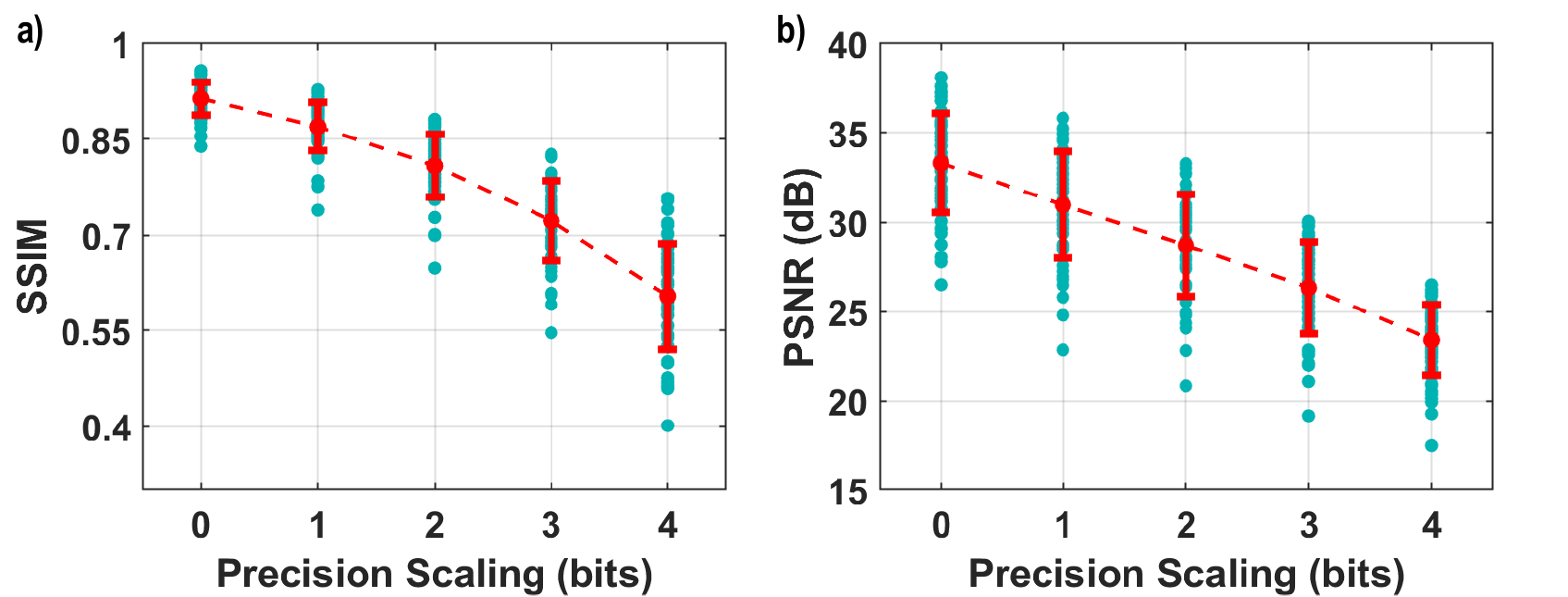}}
\caption{Effect of precision scaling: (a) SSIM vs. precision scaling level, and (b) PSNR vs. precision scaling level.}
\label{fig:precision_scaling_plot}
\vspace{-0.1in}
\end{figure}

\begin{figure}[htbp]
\centerline{\includegraphics[width=0.3\textwidth]{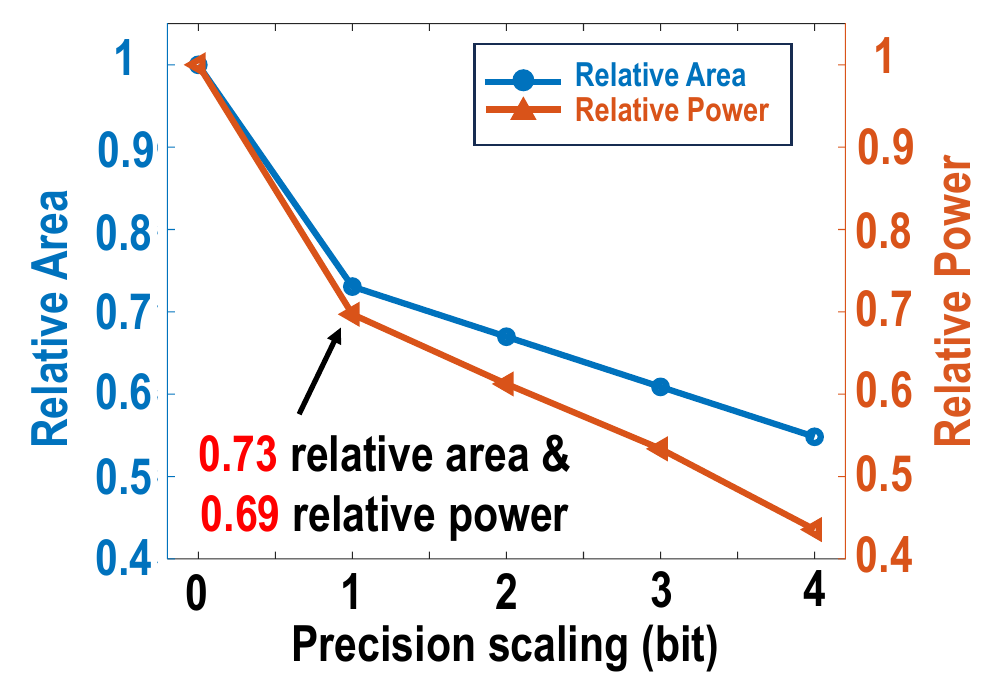}}
\caption{Hardware-level area and power improvement based on different precision scaling levels.}
\label{fig:AreaPower_BT_Q_plot}
%\vspace{-0.1in}
\end{figure}

Fig. \ref{fig:precision_scaling_plot} depicts the effect of precision scaling on the reconstructed image quality. Detailed statistics of the quality degradation is reported in Table~\ref{tab:precision_scaling_plot}. With one-bit truncation, the quality of the reconstructed images degrades slightly; only $4.82\%$ SSIM degradation and  $7.05\%$ PSNR degradation are observed according to the simulation results over the dataset.
 Hardware-wise benefit resulting from the bit truncation technique is illustrated in Fig.~\ref{fig:AreaPower_BT_Q_plot}. The JPEG compression circuit with 1-bit truncation consumes around $30\%$ less power and area than the circuit with no data bit width modification.
 
\begingroup
\begin{table*}[h]
  \centering
\caption {\label{tab:precision_scaling_plot} Statistics of Fig \ref{fig:precision_scaling_plot}: image quality under different bit truncation levels.} 
%\begin{ruledtabular}
\begin{tabular}{cccccccccc}\hline 
\multirow{2}{*}{Bits truncated} & \multicolumn{4}{c}{SSIM} &  & \multicolumn{4}{c}{PSNR}  
        \\ \cline{2-5} \cline{7-10}  
        & mean & std & max & min & & mean & std & max & min \\  \hline
$0$   & $0.9137$  & $0.0252$  & $0.9578$ & $0.8387$ &  & $33.31$  & $2.80$ & $38.11$ & $26.49$   \\       
$1$   & $0.8697$  & $0.0377$  & $0.9275$ & $0.7388$ &  & $30.96$  & $3.00$ & $35.83$ & $22.86$   \\
$2$   & $0.8084$  & $0.0487$  & $0.8814$ & $0.6477$ &  & $28.71$  & $2.86$ & $33.27$ & $20.83$   \\
$3$   & $0.7221$  & $0.0622$  & $0.8264$ & $0.5466$ &  & $28.33$  & $2.56$ & $30.05$ & $19.16$   \\
$4$   & $0.6041$  & $0.0828$  & $0.7574$ & $0.4009$ &  & $23.40$  & $1.98$ & $26.49$ & $17.49$   \\ \hline
\end{tabular}
%\end{ruledtabular}
\end{table*}
\endgroup

\begin{figure}[htbp]
\centerline{\includegraphics[width=0.4\textwidth]{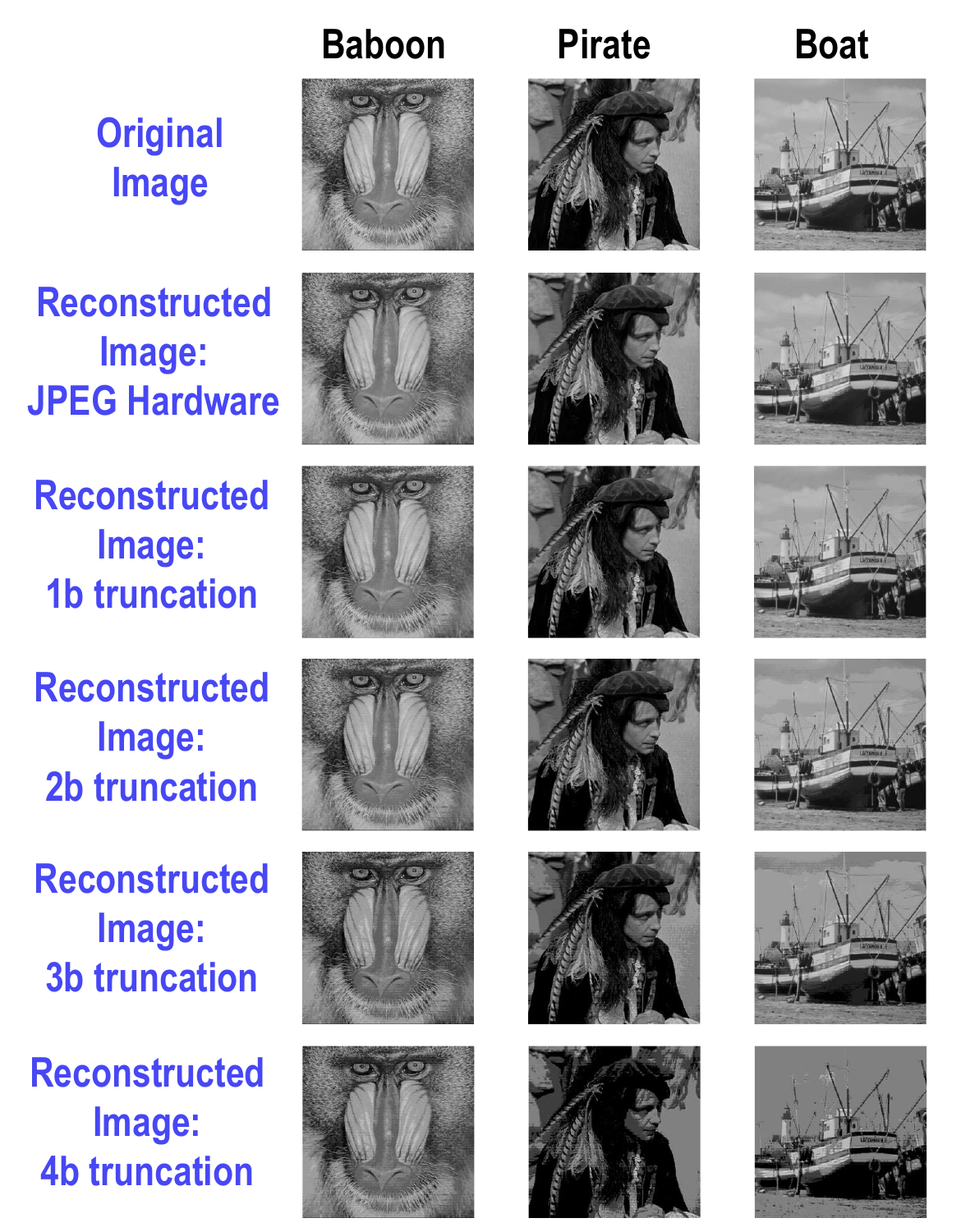}}
\caption{Reconstructed images for different precision scaling (bit truncation) levels.}
\label{fig:precision_scaling_results}
\vspace{-0.15in}
\end{figure}

Fig.~\ref{fig:precision_scaling_results} shows the subjective analysis on three selective images, Baboon, Pirate, and Boat, under different truncation levels (0 to 4). The quality of the reconstructed images degrades significantly under bit truncation levels 3 and 4. It can be seen from the images in the last two rows of Fig.~\ref{fig:precision_scaling_results} that they are heavily blurred compared to the original ones.

\subsection{Loop Perforation}

\begin{figure}[tbp]
\centerline{\includegraphics[width=0.51\textwidth]{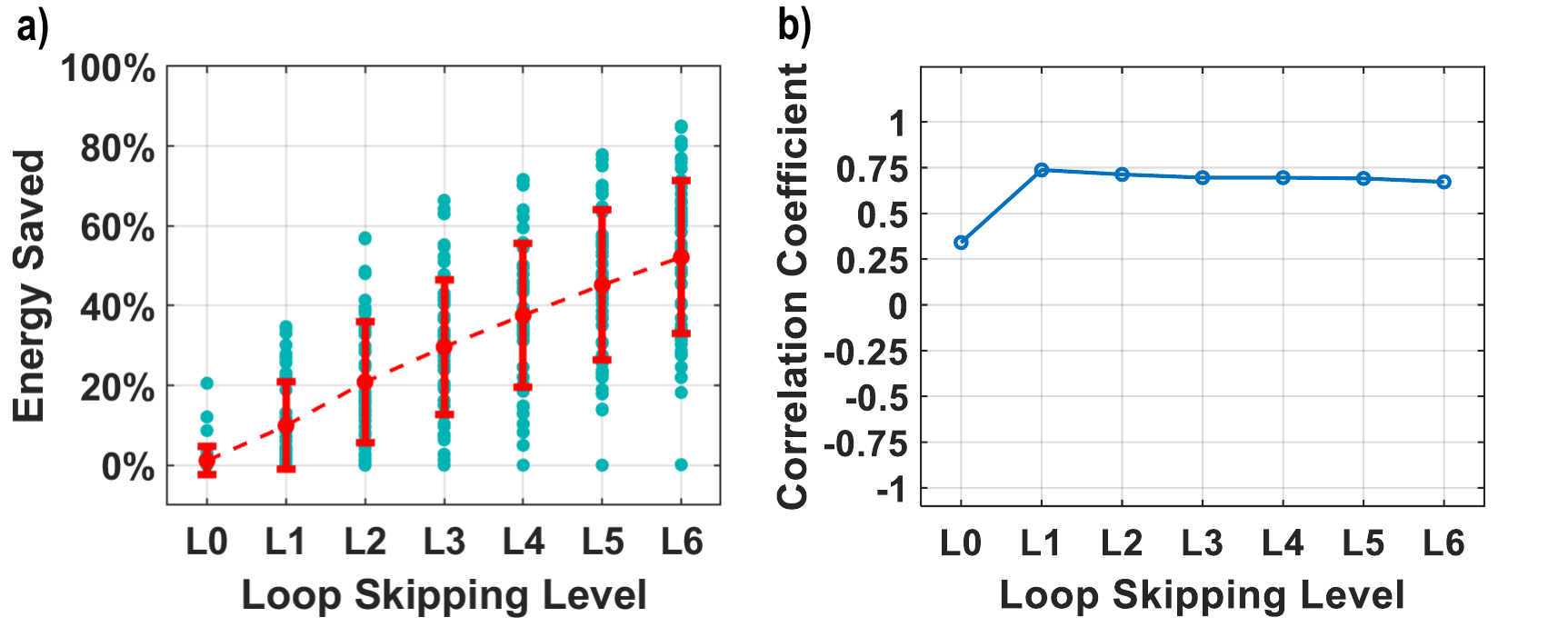}}
\caption{ Effect of different loop-skipping levels on a) energy saved and b) correlation coefficient between the energy saved and homogeneity}
\label{fig:LPSK_results_energy_saved}
%\vspace{-0.2in}
\end{figure}

Fig.~\ref{fig:LPSK_results_energy_saved}(a) and Table~\ref{tab:LPSK_results_energy_saved} present the simulation results of how much energy is saved based on a particular loop-skipping level, and Fig.~\ref{fig:LPSK_results_energy_saved}(b) shows the correlation coefficient between the energy saved and the image $homogeneity$ under different loop-skipping levels. ($Homogeneity$ is one of the Haralick textural features developed in \cite{4309314}, which is an indicator to describe pixel discrepancy in an image. Generally, the higher the homogeneity is, the more similar the neighboring pixels in an image are.) For any loop-skipping level beyond L0, a correlation coefficient larger than or close to 0.7 is observed, which implies the energy saved and the homogeneity are positively and highly correlated. Therefore, we argue that the proposed loop-skipping technique is an effective approximation scheme for signals exhibiting high spatial locality, like images, video, etc.

%\vspace{-0.1in}

\begingroup
\begin{table}[h]
  \centering
\caption {\label{tab:LPSK_results_energy_saved} Statistics of Fig \ref{fig:LPSK_results_energy_saved}: energy saved under different loop skipping levels.} 
%\begin{ruledtabular}
\begin{tabular}{ccccc}\hline 
\multirow{2}{*}{Loop-Skipping Level} & \multicolumn{4}{c}{Energy Saved}
        \\ \cline{2-5}   
        & mean & std & max & min  \\  \hline
$L_{0}$   & $1.14\;\%$  & $3.52\;\%$  & $1.14\;\%$ & $0$  \\       
$L_{1}$   & $9.9\;\%$   & $10.96\;\%$  & $34.72\;\%$ & $0$  \\
$L_{2}$   & $20.84\;\%$ & $15.14\;\%$  & $57.01\;\%$ & $0$  \\
$L_{3}$   & $29.64\;\%$ & $16.91\;\%$  & $66.41\;\%$ & $0$  \\       
$L_{4}$   & $37.57\;\%$ & $18.12\;\%$  & $71.66\;\%$ & $0$  \\
$L_{5}$   & $45.17\;\%$ & $18.82\;\%$  & $77.88\;\%$ & $0$  \\ 
$L_{6}$   & $52.14\;\%$ & $19.23\;\%$  & $85.06\;\%$ & $0.12\;\%$      \\ \hline
\end{tabular}
%\end{ruledtabular}
\end{table}
\endgroup

\par
\begin{figure}[h]
\centerline{\includegraphics[width=0.51\textwidth]{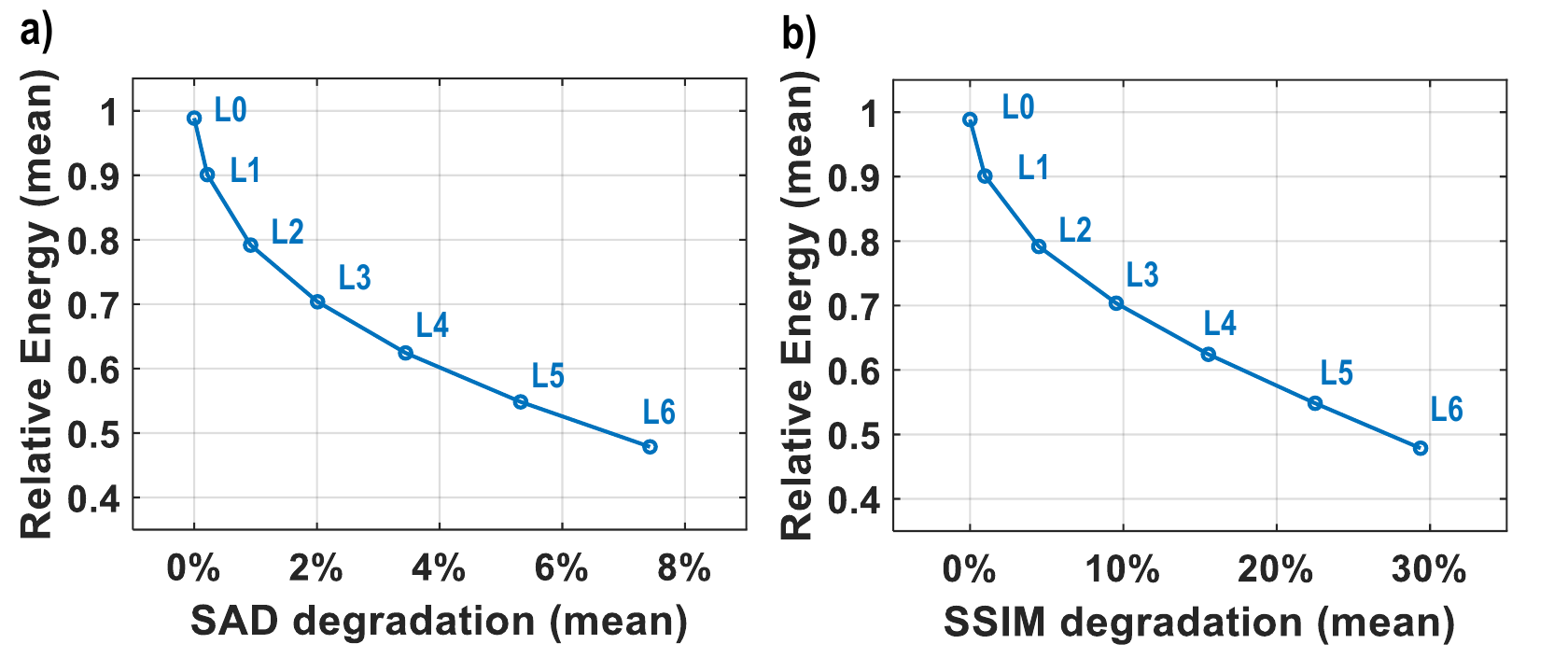}}
\caption{ Effect of different loop-skipping levels: a) relative energy vs. SAD degradation and b) relative energy vs. SSIM degradation.}
\label{fig:LPSK_degradation_24Jan}
%\vspace{-0.2in}
\end{figure}

The trade-off of the energy saved by loop skipping versus the quality degradation of reconstructed images is outlined in Fig.~\ref{fig:LPSK_degradation_24Jan}. It plots the relation between relative energy and image quality degradation for (a) SAD and (b) SSIM. The result shows that $30\%$ energy can be saved with roughly $2\%$ SAD and $10\%$ SSIM degradation at loop-skipping level $L_{3}.$

 \par
\begin{figure}[tbp]
\centerline{\includegraphics[width=0.4\textwidth]{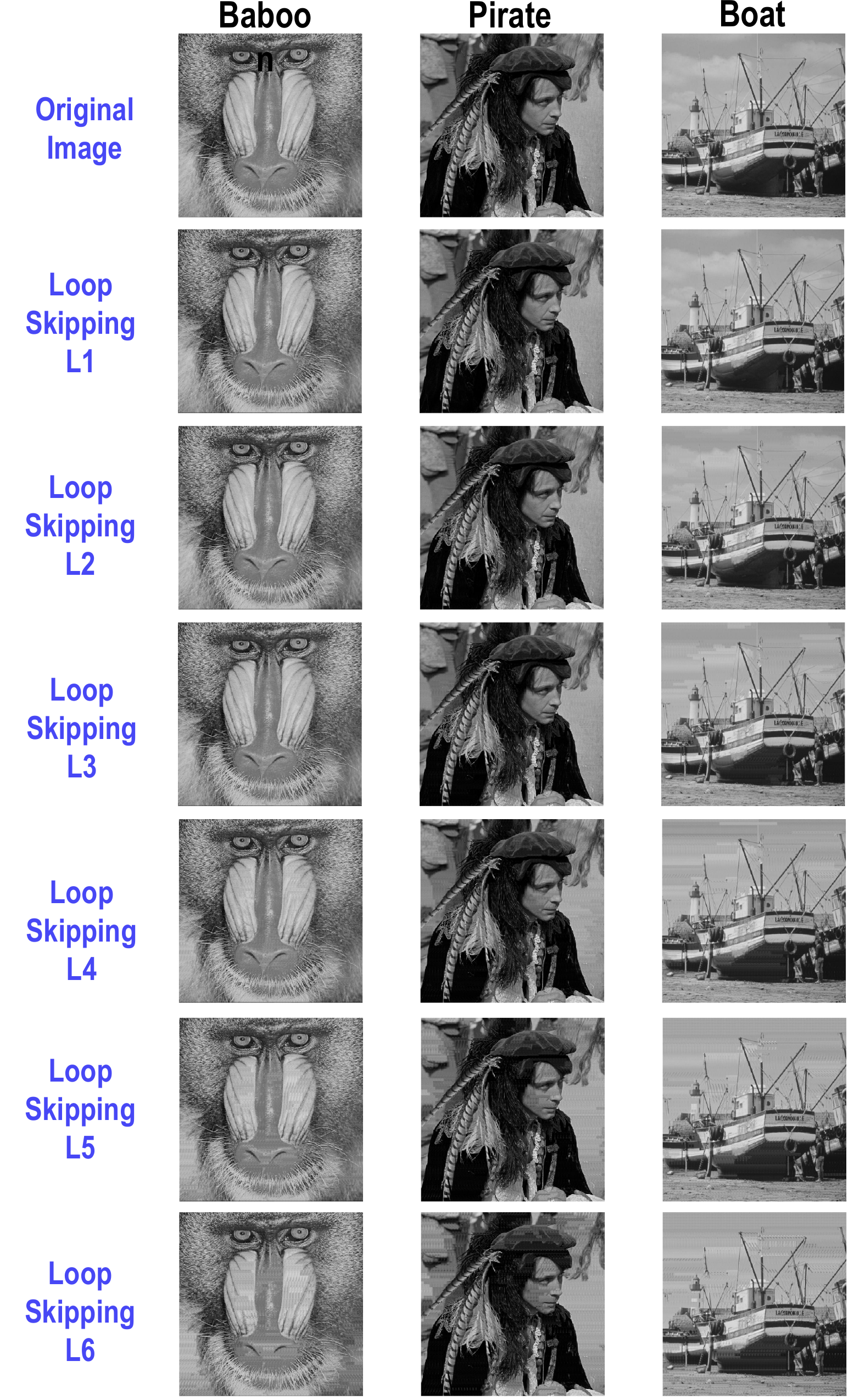}}
\caption{Effect of loop skipping: reconstructed images for different loop skipping levels.}
\label{fig:LPSK_results}
%\vspace{-0.1in}
\end{figure}

 In the end, subjective analysis of loop perforation is shown in Fig.~\ref{fig:LPSK_results}, which displays the reconstructed images of Baboon, Pirate, and Boat for different loop skipping levels.
 
\subsection{DCT}
\begingroup
\begin{table*}[h]
  \centering
\caption {\label{tab:DCT_comparison} Comparison with Existing Multiplier-less DCT Designs} 
%\begin{ruledtabular}
\begin{tabular}{ccccccc}\hline 
%\multirow{2}{*}{Item} & \multicolumn{4}{c}{Energy Saved} & d & d
%        \\ \cline{2-5}   
%        & mean & std & max & min  \\  \hline
Item & This work &  23'ASICON\cite{wang2023energy} & 21'VLSI-SoC\cite{deepsita2021energy}  & 18'ISCAS\cite{xing2018energy} & 15'ICVLSI\cite{7031760} & 15'ISVLSI\cite{darji2015high} \\
\cline{1-7}
\makecell{Approximate \\ Technique} & CSD and CSE &  \makecell{1. $\mathbf{T}$ coefficients approx. \\ 2. Optimized DCT by \\ using adjacent \\ pixel correlation} & \makecell{Approx. Integer DCT} & \makecell{1. $\mathbf{T}$ coefficients approx. \\ 2. Approx. adders} & \makecell{An approx. DCT for \\ only 25 coefficients} & \makecell{CSD and CSE} \\
\cline{1-7}
Technology   & 65 nm  & 28 nm  & 45 nm & 180 nm & 45 nm & 90 nm\\       
Area ($um^{2}$)   & $6,500$  & $12,687$   & $3,217$  & $876k$ & $3,706$ & $27,870$  \\
%Gate count   & 3k & N.A. & N.A. & N.A. & N.A. & N.A. & 149k  \\
Power ($mW$)   & $1.64$ & $19.86$  & $0.77$   & $10.34$ & $0.45$ & $2.3$ \\
Energy ($pJ$) & $16.4$ & $198$ & $14.6$ & $77.9$ & $29.2$ & $23$\\
\hline
\end{tabular}
%\end{ruledtabular}
\end{table*}
\endgroup

Several multiplier-less DCT architectures have been proposed these years. For example, \cite{wang2023energy} proposed an efficient 1D-DCT structure that approximated the coefficients of the DCT transform matrix $\mathbf{T}$ as the power of 2 and exploited adjacent pixel correlation to reduce hardware complexity. Similarly, the idea of approximating $\mathbf{T}$'s coefficients as the power of 2 was applied in \cite{deepsita2021energy} to realize an approximate integer DCT scheme for HEVC. Other work such as \cite{xing2018energy}, utilized approximate adders to build an energy-efficient DCT. \cite{7031760} proposed a 25-coefficient 
DCT architecture by exploiting pixels' correlation and relative significance of DCT coefficients. \par
We compare the hardware simulation results of this work's DCT scheme, which uses the techniques of CSD (Canonic Signed Digit) and CSE (Common Subexpression Elimination) in \cite{5156873}, with other existing 8-point multiplier-less DCT works in Table. \ref{tab:DCT_comparison}. The 1D-DCT structure utilized in this paper is described in Verilog and synthesized by Synopsys Design Compiler with TSMC 65nm process. The circuit's power performance is measured by Cadence Spectre Simulator, where 256 8-by-1 column vectors of random pixels are sent as inputs to the circuit, and the average power consumption is measured under 1V supply and 100MHz clock. The final result shows that our 1D-DCT structure consumes 16.4pJ, which is better than most of the existing works.

%  Canonical Signed Digit \\ Common Sub-expression \\ Elimination

\subsection{Results from final architecture: quality, area \& energy}\label{sec:result_all}

The final JPEG compression circuit is synthesized using the Synopsys Design Compiler tool and mapped to TSMC 65nm library. The synthesis result shows that the final architecture (with the proposed bit-shifting-based quantization block and the loop skipping function) occupies a cell area of $109,470$ ${um}^{2}$, which is 28\% less than the baseline (with conventional division-based quantization block and without the loop skipping function) design ($151,617$ ${um}^{2}$). 

\par
\begin{figure}[htbp]
\centerline{\includegraphics[width=0.51\textwidth]{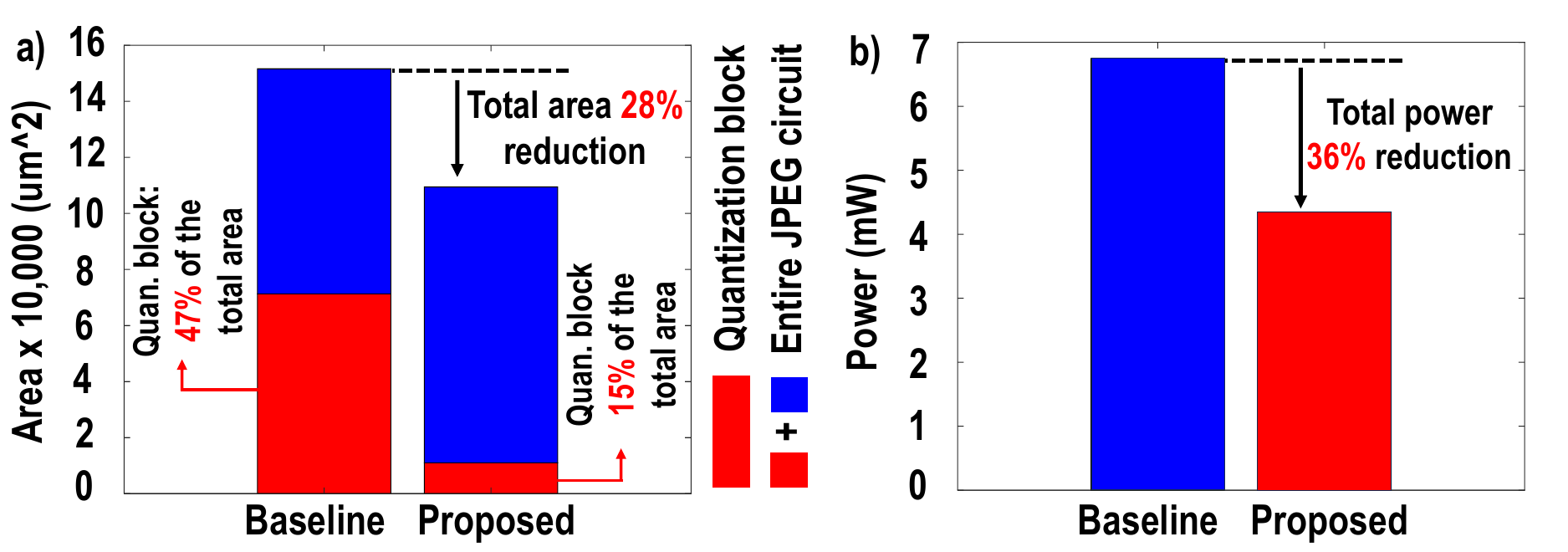}}
\caption{Comparison between the baseline and proposed design: a) Area breakdown. b) Power consumption.}
\label{fig:area_results}
\vspace{0.1in}
\end{figure}

 The area reduction comes mainly from improving the quantization step through the bit shift operators-based division, even though, at the same time, bit-truncation and loop skipping induce some area overhead. The baseline quantization blocks, comprising synthesis tool-generated dividers, consume 47\% ($71,285$ ${um}^{2}$) of the entire area; however, the optimized quantization takes only 15\% ($11,013$ ${um}^{2}$) of the total area. This area saving, $71,285-11,013=60,272$ ${um}^{2}$, outweighs the additional area resulting from the logic and registers used for bit-truncation and loop skipping, which takes $(109,470-11,013)-(151,617-71,285)=98,457-80,332=18,125$ ${um}^{2}$. Our proposed methods combined, therefore, give an area saving of $42,147$ ${um}^{2}$. Fig.~\ref{fig:area_results}(a) shows the area comparison of the baseline and proposed design, where the red bars represent the area overhead of the quantization blocks and the blue bars represent the non-quantization part.

\par According to the simulation of Cadence Spectre Simulator, at TT corner, $25^{\circ}$C, with a supply of 1V, the baseline design (using bit truncation level 1 and loop skipping level 2) consumes an average current of 6.75 mA at 100MHz at the expense of $2\%$ SAD degradation. On the other hand, the proposed architecture consumes an average current of 4.35 mA under the same simulation condition. Therefore, $36\%$ energy is saved from our proposed approach, as shown in Fig.~\ref{fig:area_results}(b). The proposed architecture equivalently dissipates a power of 15uW in the DCT and quantization stages to generate a throughput of 480p colored image @ 6fps. This is $10\times$ better than the analog solution \cite{SWCapMJPEG} (which utilizes passive elements, i.e., switch capacitors to save power) and $6\times$ better than current state-of-the-art \cite{isscc_14} (which operates at near-threshold to reduce power).

\begin{comment}
  \begingroup
\begin{table}[h]
  \centering
\caption {\label{tab:LPSK_results_energy_saved} Fig \ref{fig:LPSK_results_energy_saved}.} 
%\begin{ruledtabular}
\begin{tabular}{ccccc}\hline 
d & \makecell{Approximate \\ Technique} & Technology & Area & Power \\
\cline{1-5}
15'ISVLSI\cite{darji2015high} & \makecell{CSD and CSE} & & & a \\
15'ICVLSI\cite{7031760} & \makecell{1. $\mathbf{T}$ coefficients approx. \\ 2. Approx. adders} & & & a\\
18'ISCAS\cite{xing2018energy} & & & & a\\
21'VLSI-SoC\cite{deepsita2021energy} & & & & a\\
23'ASICON\cite{wang2023energy} & & & & a\\
\cline{1-5}
This work & & & & a\\
\hline
\end{tabular}
%\end{ruledtabular}
\end{table}
\endgroup  
\end{comment}

\section{Discussion}\label{sec:Discussion}
This work focuses on accelerating the DCT and quantization steps in JPEG compression. However, another widely used image compression scheme, JPEG2000, merits discussion due to its superior compression efficiency and image quality.

\subsubsection{ JPEG vs. JPEG2000 }
Although the transform step in JPEG2000, which uses discrete wavelet transform (DWT), is simpler than the discrete cosine transform (DCT) in JPEG, the overall structure of JPEG2000 is more complex\cite{Liang-Gee_Chen_Analysis_and_architecture_design_of_block-coding_engine_for_EBCOT_in_JPEG_2000} \cite{adams2002jpeg} \cite{chen2001analysis} \cite{Design_Methodology_of_Low_Power_JPEG2000_Codec_Exploiting_Dual_Voltage_Scaling}\cite{JPEG2000_Hardware_Implementation_Procedures_and_Issues}\cite{A_VLSI_architecture_of_JPEG2000_encoder}. This complexity arises from more refined quantization steps involving floating-point division and advanced coding schemes like Embedded Block Coding with Optimized Truncation (EBCOT) \cite{Taubman_High_performance_scalable_image_compression_with_EBCOT}. Consequently, JPEG2000 is not well-suited for applications in edge devices.
\subsubsection{ DCT vs. DWT}
The 2D-DCT is more complex than the 2D-DWT for the transformation step. Nonetheless, the multiplier-less approach adopted in this design significantly reduces the hardware implementation cost, making it competitive with the 2D-DWT in terms of hardware resources. The transform block in the proposed design performs the 2D-DCT using only shift, addition, and subtraction operations—no multipliers are used. This approach is highly similar to the implementation of FIR-based 2D-DWT.
\subsubsection{ Possible future work for approximate JPEG2000 }
 The approximate techniques proposed in this paper—multiplier-less transformation, approximate quantization, bit truncation, and loop perforation—can be applied not only to JPEG but also to JPEG2000. For instance, 2D-DWT can be implemented using only shift, addition, and subtraction operations, as it is essentially an FIR filter. Quantization in JPEG2000 can also be implemented through shifting logic. Additionally, since JPEG2000 compresses images in small tiles, the concept of loop skipping can be utilized. If two neighboring tiles are sufficiently similar, the compression unit can be disabled, and the results from the last processed tile can be reused. Our future work can include detailed approximated hardware analysis and implementation and related encoded/decoded image analysis for JPEG2000.

%\textcolor{red}{KGK provide final baseline power}. 
\section{Conclusion}\label{sec:final}
This work demonstrates a synthesizable multiplier-less JPEG accelerator equipped with approximations both in software and RTL in the form of modified quantization block, precision scaling, and loop perforation, trading off the quality of the image with energy \& area reduction. With a gradient descent-based heuristic, the accelerator's performance can be tuned to maximize energy savings while meeting the image quality constraints. 
The proposed architecture with the combined approximation strategies achieves $36\%$ reduction in energy consumption at the expense of $2\%$ SAD quality degradation in the image, which lies within acceptable limits for any image processing applications. Moreover, it consumes 15uW at the DCT and quantization stages to compress a colored 480p image at 6fps, which is 10x better than the previous literature. 
%NEW citations: \cite{The_JPEG_2000_still_image_compression_standard} \cite{The_JPEG2000_still_image_coding_system_an_overview}

\normalem
\bibliographystyle{unsrt} % We choose the "plain" reference style
\bibliography{references.bib} % Entries are in the references.bib file in bibtex format
\ULforem

\begin{IEEEbiography}[{\includegraphics[width=1in,height=1.25in,clip,keepaspectratio]{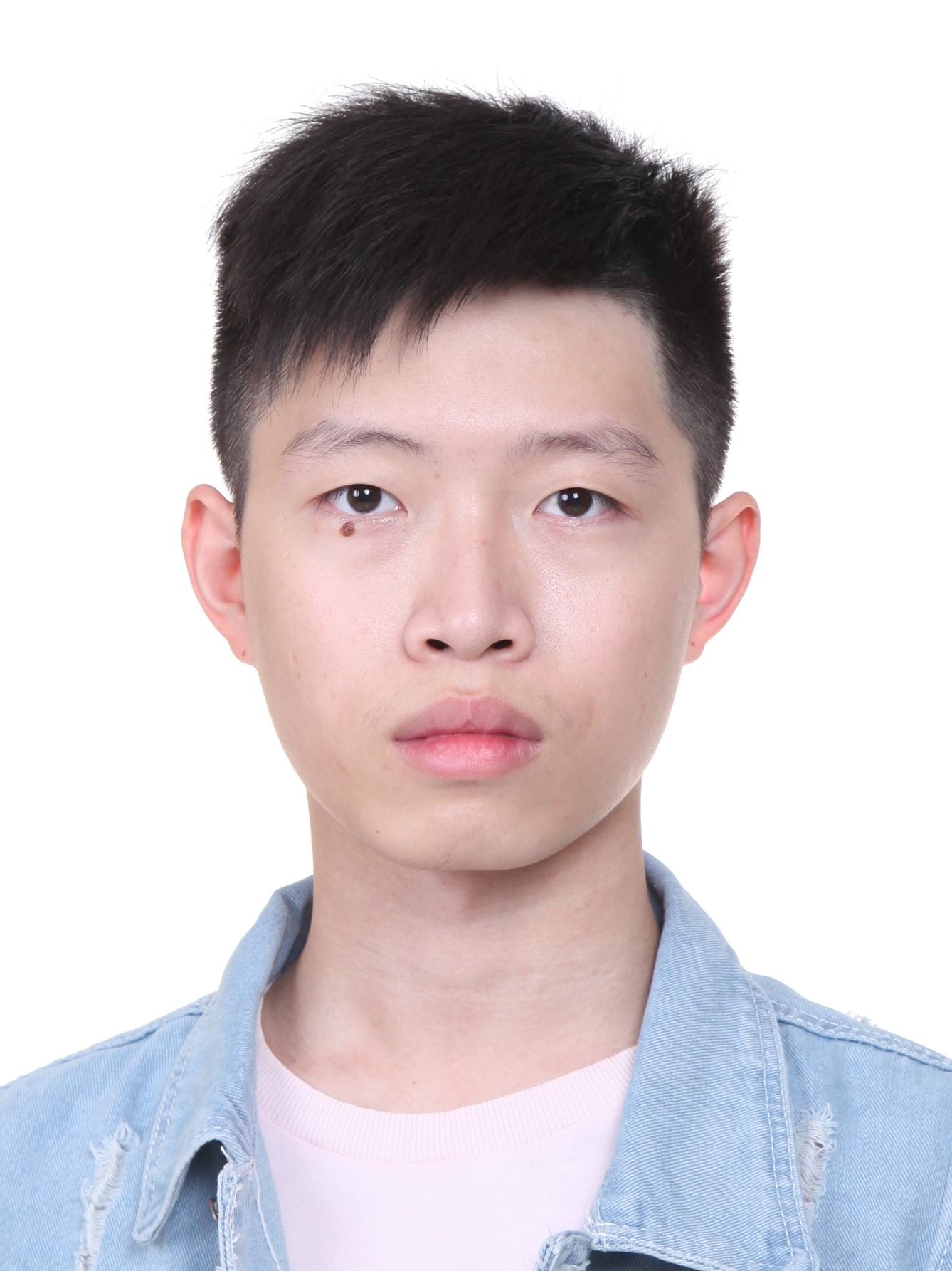}}]{Ming-Che Li} (Graduate Student Member, IEEE) was born in Yilan, Taiwan, in 1999. He received the B.S. degree in electrical engineering from National Tsing Hua University (NTHU), Hsinchu, Taiwan, in 2021. He is currently pursuing a Ph.D. degree in electrical and computer engineering at Purdue University, West Lafayette, IN, USA.\par His current research interests include hardware security, approximate computing in image \& video compression, and stochastic computing.
\end{IEEEbiography}

\begin{IEEEbiography}[{\includegraphics[width=1in,height=1.25in,clip,keepaspectratio]{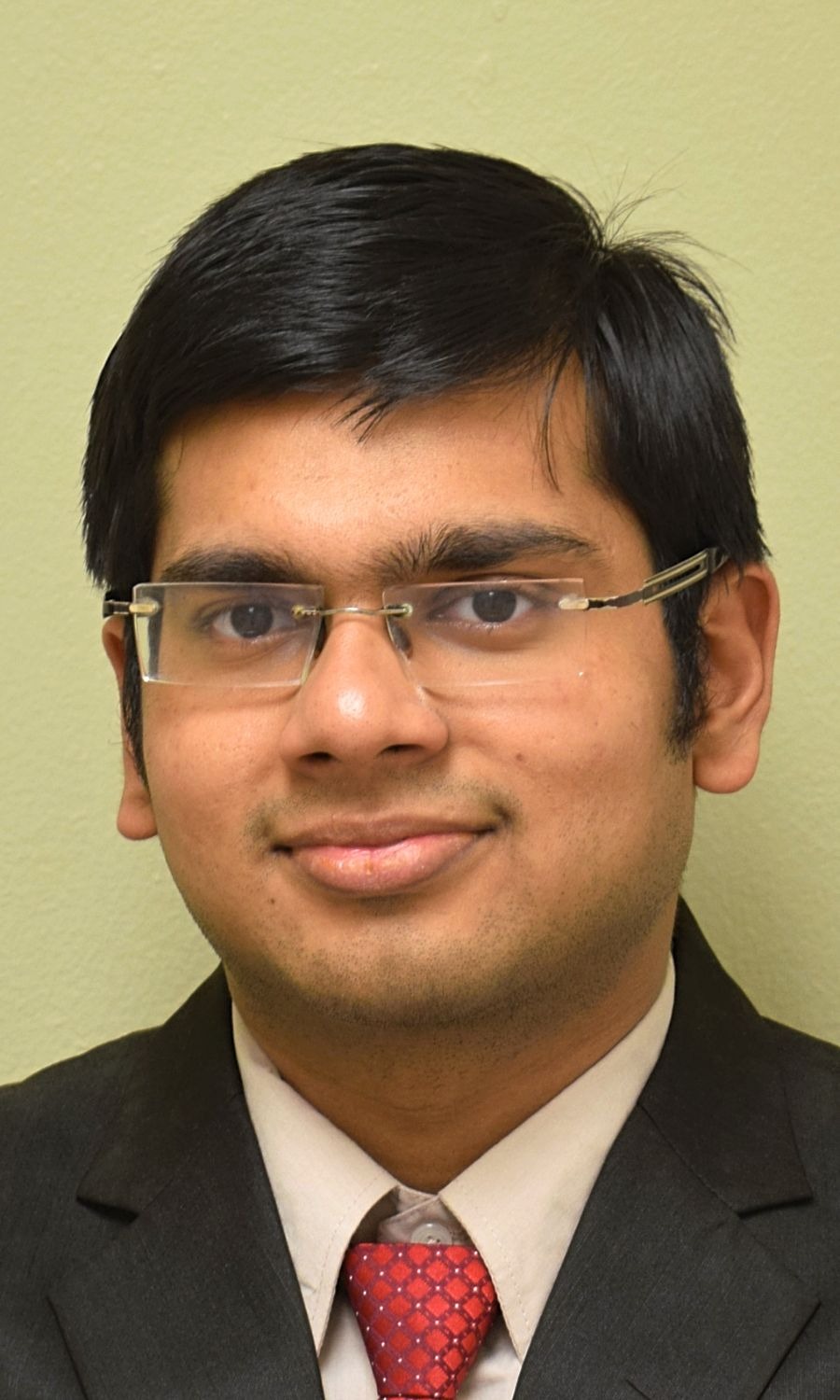}}] {Archisman Ghosh} (Student Member, IEEE) received his B.E. degree in Electronics and Telecommunication Engineering from Jadavpur University, India, in 2017, and he is currently pursuing a Ph.D. at Purdue University, where he was a recipient of prestigious ECE Meissner fellowship (2019-20) as an incoming graduate student. He is currently a Bilsland Dissertation fellow at Purdue University. \par
His research interests include digital SoC design and hardware security. Prior to his Ph.D., Mr. Ghosh worked in Samsung Semiconductor India R\&D for 2 years. He has interned with Intel Labs, Oregon. He is one of the recipients of the prestigious IEEE SSCS Pre-doctoral Achievement Award 2022.
\end{IEEEbiography}

\begin{IEEEbiography}[{\includegraphics[width=1in,height=1.25in,clip,keepaspectratio]{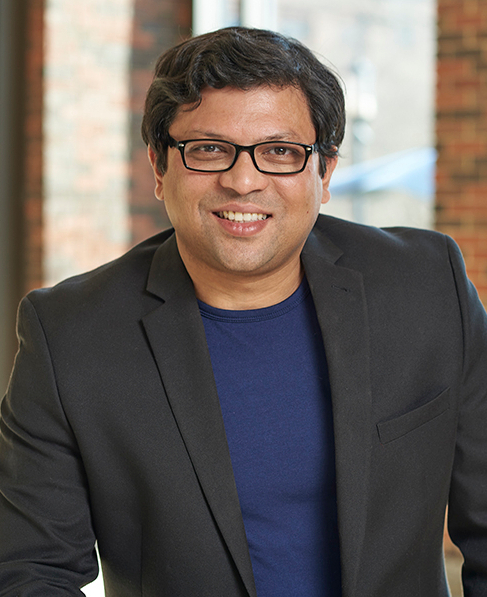}}]{Shreyas Sen}
Shreyas Sen is an Elmore Associate Professor of ECE \& BME, Purdue University. His current research interests span mixed-signal circuits/systems and electromagnetics for the Internet of Bodies (IoB) and Hardware Security. He has authored/co-authored 3 book chapters, over 200 journal and conference paper and has 25 patents granted/pending. Dr. Sen serves as the Director of the Center for Internet of Bodies (C-IoB) at Purdue. Dr. Sen is the inventor of the Electro-Quasistatic Human Body Communication (EQS-HBC), or Body as a Wire technology, for which, he is the recipient of the MIT Technology Review top-10 Indian Inventor Worldwide under 35 (MIT TR35 India) Award in 2018 and Georgia Tech 40 Under 40 Award in 2022. To commercialize this invention Dr. Sen founded Ixana and serves as the Chairman and CTO and led Ixana to awards such as 2x CES Innovation Award 2024, EE Times Silicon 100, Indiana Startup of the Year Mira Award 2023. His work has been covered by 250+ news releases worldwide, invited appearance on TEDx Indianapolis, NASDAQ live Trade Talks at CES 2023, Indian National Television CNBC TV18 Young Turks Program, NPR subsidiary Lakeshore Public Radio and the CyberWire podcast. Dr. Sen is a recipient of the NSF CAREER Award 2020, AFOSR Young Investigator Award 2016, NSF CISE CRII Award 2017, Intel Outstanding Researcher Award 2020, Google Faculty Research Award 2017, Purdue CoE Early Career Research Award 2021, Intel Labs Quality Award 2012 for industrywide impact on USB-C type, Intel Ph.D. Fellowship 2010, IEEE Microwave Fellowship 2008, GSRC Margarida Jacome Best Research Award 2007, and nine best paper awards including IEEE CICC 2019, 2021 and in IEEE HOST 2017-2020, for four consecutive years. Dr. Sen's work was chosen as one of the top-10 papers in the Hardware Security field (TopPicks 2019). He serves/has served as an Associate Editor for IEEE Solid-State Circuits Letters (SSC-L), Nature Scientific Reports, Frontiers in Electronics, IEEE Design \& Test, Executive Committee member of IEEE Central Indiana Section and Technical Program Committee member of TPC member of ISSCC, CICC, DAC, CCS, IMS, DATE, ISLPED, ICCAD, ITC, and VLSI Design. Dr. Sen is a Senior Member of IEEE.
\end{IEEEbiography}

\end{document}